\documentclass[universe,article,accept,moreauthors,pdftex]{Definitions/mdpi}

\firstpage{1}
\pubvolume{7}
\issuenum{3}
\articlenumber{76}
\pubyear{2021}
\copyrightyear{2021}
\history{Received: 24 February 2021; Accepted: 11 March 2021; Published: 23 March 2021}

\usepackage{amssymb}
\usepackage{Definitions/aas_macros}
\Title{Multiwavelength Observations of Fast Radio Bursts} %Attention AE/ME. The following layout issues have not been checked by the English Editing Department and must be carefully verified by the AE/Layout Department: All callout issues, bold usage of callouts, and references to callouts in the text. Correct callout usage in figures. Figure and Table layout issues. Footnote formatting and Glossaries have not been checked. En dash usage for negative values, en dash usage to indicate relationships, en dash usage to indicate bonds (especially in chemistry). The English Editing Department is not responsible for correct italic usage for genes, proteins and technical terminology. This responsibility belongs to the authors. The following are also not checked: spacing between numbers and units of measurement, ratios, en dashes for ranges, date and time formats, punctuation in equation lines, and less than/more than spacing (< >). Finally, capitalization and layout of titles/headings must be properly checked as well as ensuring 'Eq.' and 'Fig.' are properly spelled out, as these are layout issues.

\newcommand{\fdeg}{$\,.\!\!^{\circ}$}
\newcommand{\farcm}{$\,.\!\!^{\prime}$}
\newcommand{\farcs}{$\,.\!\!^{\prime\prime}$}

\newcommand{\dataavailability}[1]{%
\vspace{12pt}\noindent{\fontsize{9}{9}\selectfont\textbf{Data Availability Statement:} {#1}\par}}

\Author{Luciano  Nicastro $^{1,}$*\href{https://orcid.org/0000-0001-8534-6788}{\orcidicon}, Cristiano Guidorzi $^{1,2,3}$\href{https://orcid.org/0000-0001-6869-0835}{\orcidicon}, Eliana Palazzi $^{1}$\href{https://orcid.org/0000-0002-8691-7666}{\orcidicon}, Luca Zampieri $^{4}$\href{https://orcid.org/0000-0002-6516-1329}{\orcidicon}, Massimo Turatto $^{4}$\href{https://orcid.org/0000-0002-9719-3157}{\orcidicon} and Angela Gardini $^{5}$\href{https://orcid.org/0000-0002-3234-9130}{\orcidicon}}
\address{%
$^{1}$ \quad INAF---Osservatorio di Astrofisica e Scienza dello Spazio di Bologna, Via Piero Gobetti 93/3, I-40129 Bologna, Italy; guidorzi@fe.infn.it (C.G.); eliana.palazzi@inaf.it (E.P.) \\
$^{2}$ \quad Department of Physics and Earth Science, University of Ferrara, Via Saragat 1, I-44122, Ferrara, Italy\\
$^{3}$ \quad INFN---Sezione di Ferrara, Via Saragat 1, I-44122 Ferrara, Italy\\
$^{4}$ \quad INAF---Osservatorio Astronomico di Padova, Vicolo dell'Osservatorio 5, I-35122 Padova, Italy; luca.zampieri@inaf.it (L.Z.); massimo.turatto@inaf.it (M.T.)\\
$^{5}$ \quad Observatorio Astron\'omico Calar Alto s/n, Sierra de los Filabres, E-04550 Gergal, Almería, Spain; gardini@caha.es
}

\corres{Correspondence: luciano.nicastro@inaf.it; Tel.: +39-051-6398778}

\abstract{
The origin and phenomenology of the Fast Radio Burst (FRB) remains unknown despite more than a decade of efforts. Though several models have been proposed to explain the observed data, none is able to explain alone the variety of events so far recorded. The leading models consider magnetars as potential FRB sources. The recent detection of FRBs from the galactic magnetar SGR~J1935+2154 seems to support them. Still, emission duration and energetic budget challenge all these models.
Like for other classes of objects initially detected in a single band, it appeared clear that any solution to the FRB enigma could only come from a coordinated observational and theoretical effort in an as wide as possible energy band. In particular, the detection and localisation of optical/NIR or/and high-energy counterparts seemed an unavoidable starting point that could shed light on the FRB physics.
Multiwavelength (MWL) search campaigns were conducted for several FRBs, in particular for repeaters. Here we summarize the observational and theoretical results and the perspectives in view of the several new sources accurately localised that will likely be identified by various radio facilities worldwide.
We conclude that more dedicated MWL campaigns sensitive to the millisecond--minute timescale transients are needed to address the various aspects involved in the identification of FRB counterparts. Dedicated instrumentation could be one of the key points in this respect. In the optical/NIR band, fast photometry looks to be the only viable strategy. Additionally, small/medium size radiotelescopes co-pointing higher energies telescopes look a very interesting and cheap complementary observational strategy.
}


\begin{document}
%
%%%%%%%%%%%%%%%%%%%%%%%%%%%%%%%%%%%%%%%%%%
\section{Introduction}
The multiwavelength (MWL) approach to study transient astronomical events has demonstrated its effectiveness in solving many puzzles in astronomy, both related to ``local'' and extragalactic sources. The Gamma Ray Burst (GRB) phenomenon represents a perfect example. Wide area detectors (e.g., high-energy monitors) or specific surveys monitoring large areas of the sky can detect events that only last for a short period of time. Being able to reduce positional uncertainties and perform timely MWL observations using sensitive and high-enough resolution instruments could be the only way to discriminate among the possible progenitors. This is a key point when the transient astronomical event is only detected in a single band of the electromagnetic spectrum or when multiple sources could produce that event. Studying the source emission in a as wide as possible spectral band justifies huge efforts in terms of observational time and manpower. It, typically, turns out to be the only way to solve the most challenging questions in astrophysics.

Fast Radio Bursts (FRBs) are among the most studied astrophysical transients, still their origin and whether there are multiple types of progenitors and emission mechanisms are still open questions
(see~\cite{Katz18rev, Petroff19rev, Cordes19rev, Zhang20Nat, Xiao21rev} for a review).
%R: Moved here from the Abstract section
Are ({\em apparently}\/) one-off and repeating events representative of distinct samples or are they the realization of a very wide timescales distribution of the same objects class?
Can the periodicity seen in a few FRBs be reconciled with the proposed models? Is the behavior distinctive of a class or sub-class of FRBs?
\textls[-15]{Of the about 140 distinct sources {known} [\url{https://wis-tns.weizmann.ac.il/}] (accessed on 1~March~2021) (but we are aware that many more have been discovered but remain unpublished at the time of writing this review), only for a bunch of them a MWL search campaign was possible, in particular for the two repeating (and periodic) sources FRB~20121102A (commonly referred as FRB~121102) and FRB~20180916B (also referred as FRB~180916 and FRB~180916.J0158+65, see below for the details).
%please confirm if add accessed date for website
%R: we prefer not to add dates
\textls[-25]{What makes FRB searches even more challenging than for other transients is the duration of the event (before its flux falls below our detection limit) at ``all'' wavelengths. For example while a short GRB detected in the $\gamma$-rays can also last a few tens of milliseconds~\cite{Cline99, Cline05}, it can remain detectable at other wavelengths for days or longer. In the FRB case, even though some of the many, still viable, emission mechanisms predict a sort of afterglow emission similar to that
of GRBs, they also predict a very weak signal on time scales of (at most) minutes after the radio burst. Therefore it seems much more promising searching for an almost simultaneous, ms-duration burst also at wavelengths outside the radio band. The recently proposed unified magnetar models by Lu et al. (2020)~\cite{Lu20} and \mbox{Margalit et al. (2020a)~\cite{Margalit20uni}} support this scenario. This would then require MWL simultaneous observational campaigns and the use of detectors capable of acquiring data at a high cadence. Fine time resolution is normal for high-energy detectors on-board satellites, much less for on-ground \mbox{optical/NIR cameras.}}}

\textls[-15]{In spite of the lack of MWL detections, possibly due to the limited capabilities of existing instruments, there is no doubt that also non-detections in FRB follow-up campaigns remain of great importance.
Collecting observational data and flux upper limits at all wavelengths are helpful to constrain rate and spectral properties, as well as to identify periods of active emission phases and then estimate the probability of events detection. This is the case for repeating FRBs (rFRBs). Upper limits on fluxes are also relevant to constrain the fluence ratios between the high energy bands (optical and X-/$\gamma$-ray) and the radio band, which in turn put constraints on the proposed FRB emission models (e.g.,~\cite{Platts19}, \url{http://frbtheorycat.org} (accessed on 1 March 2021)). At the current stage of the FRB research, observational data that can rule out theories represent a highly valuable work.
Vacuum synchrotron maser, plasma synchrotron maser and synchrotron maser from magnetized shocks, coherent curvature emission, are among the most invoked mechanisms (see e.g.,~\cite{Zhang20Nat, Xiao21rev, Lyubarsky21} for a review) but, as it was the case for GRBs, the controversy on which radiation mechanism fits best the data may last awhile before reaching a final conclusion. With the additional complication of the (apparent/real) dicothomy of one-off and repeating bursts.
Meanwhile, coordinated MWL observing campaigns, in particular of rFRBs, represent a key point to verify/challenge \mbox{their predictions}.}

As new MWL observational data are being published on the transient emission from the Galactic magnetars SGR~J1935+2154~\cite{Chime20SGR1915, Bochenek20Nat, Zhang20_FAST1935, Mereghetti20, Lin20b, LiHXMT20, Ridnaia20}, 1E~1547.0--5408~\cite{Israel21}, XTE~J1810--197~\cite{Maan19}, Swift~J1818.0--1607~\cite{Esposito20}, similarities with the FRB phenomena become more and more striking, and then the possible common physical processes involved~\cite{Lu20}.
On the other hand MWL campaigns on FRB~20180916B can rule out the occurrence of magnetar giant flares (MGF) ($E < 10^{45-47}$~erg) either simultaneous to a few radio bursts, or in general during some of the radio-burst active phases~\cite{Scholz20, Pilia20, Tavani20, Casentini20, Guidorzi20b} and constrain the possible associated persistent X-ray luminosity to <$2 \times 10^{40}$~erg~s$^{-1}$~\cite{Scholz20}, which is still decades above the observed persistent luminosity of magnetars.
In addition, the possible existence of a population of extragalactic magnetars that are equally or even more active than their Galactic siblings and that can emit even more energetic flares~\cite{Mazets08, Burns21}, as was also the recent case of a MGF from NGC~253 (Sculptor Galaxy) at 3.5~Mpc~\cite{Yang20, Svinkin21, Roberts21}, adds to the case of monitoring the high-energy activity of nearby rFRBs.
In parallel, the expected growing sample of rFRBs in the coming years will enable a systematic search for past activity hidden in the optical and high-energy surveys, as was done for the few known cases (e.g.,~\cite{Yamasaki16, ZhangZhang17, Yang19, Sun19, Andreoni20}).

Regarding the FRBs host galaxies, as of today 13 have been firmly identified. Such limited sample does not yet allow us to draw solid conclusions about potential progenitors as observational selection biases could play an important role. However statistical studies of stellar masses and star formation rates (SFRs) suggest that, at least some of them, are consistent with the host galaxies of core-collapse supernovae (CCSNe), but not with the hosts of long GRBs (LGRBs) and superluminous supernovae (SLSNe-I)~\cite{Heintz20, Bochenek21mag, Mannings20}.
This strengthens the possibility that FRBs are produced by magnetars. As a larger sample of FRB hosts becomes available, possibly with offset distribution and local environment studies, it may turn up evidence for alternate magnetar formation channels or call for a second progenitor scenario for FRBs.

{FRB~20121102A~\cite{Spitler14} was the first FRB for which multiple bursts were detected, and is then known as the ``repeating FRB''~\cite{Spitler16}.
Karl G. Jansky Very Large Array (VLA) sub-arcsec localisation allowed its host galaxy at $z \simeq 0.193$ to be identified~\cite{Chatterjee17, Marcote17, Tendulkar17}.
FRB~20180916B~\cite{chime19_8repeaters} was discovered by the Canadian Hydrogen Intensity Mapping Experiment (CHIME) and was immediately identified as a repeater. Follow-up very long baseline interferometry (VLBI) campaigns led to its precise localisation and the identification of the host galaxy at a redshift $z \simeq 0.0337$~\cite{Marcote20}.
This identification, second ever for a rFRB, immediately showed a dichotomy with the case of the original repeater, with FRB~20180916B associated to a star-forming region within a nearby massive spiral galaxy whereas FRB~20121102A host is a low-metallicity dwarf galaxy.
The subsequent continuous monitoring of FRB~20180916B by CHIME led to the first identification of a periodicity in the active phases of a rFRB~\cite{Chime20-180916p}, recurring every $16.3$~days and with an active window phase of approx $\pm 2.6$~days around the midpoint of the window.
Thanks to the continuing monitoring and bursts collection, a periodicity of $161 \pm 5$ days in the FRB~20121102A bursts was later claimed by~\cite{Rajwade20, Cruces21}. Models to explain this recurring active phases are growing, with the most recent one invoking a potential connection to ultra-luminous X-ray sources (ULXs), the closest known persistent super-Eddington sources~\cite{Sridhar21}.
More about these two peculiar FRBs in the \mbox{Section \ref{sec:R1_R3}.}}

\textls[-15]{In this paper we review the outcome of most FRB MWL searches reported in the literature, discuss the capabilities of present and being built instrumentation and what we believe are the most promising strategies to adopt in future campaigns.
In Section \ref{sec:magnetars} we introduce magnetars and the FRB~20200428A detected from SGR~J1935+2154. We discuss the characteristics of the currently identified FRB host galaxies in Section \ref{sec:Hosts}. A critical comparison of the various transient source hosts is also presented.
In Section \ref{sec:FRBs_MWLsearches} we illustrate the various efforts and outcome from the observational campaigns and archival searches for the high-energy counterpart of FRBs, from the optical band to the very high-energy (VHE) $\gamma$-rays. We focus in particular on coordinated observational campaigns, being the most promising approach in light of the (quasi-)simultaneous MWL emission predicted by the magnetar-engine models.
The most favoured emission models are also briefly discussed.
FRBs $\gamma$-ray energetic is compared to the radio one and to that of GRBs and galactic magnetars.
In Section \ref{sec:R1_R3} optical and higher-energy observations of the two periodic repeaters FRB~20121102A and FRB~20180916B are extensively discussed.
The recent outcome from the MWL observations performed during the April 2020 SGR~J1935+2154 active phase are illustrated in Section \ref{sec:SGR1935}. Our conclusions are summarized in Section \ref{sec:conclusions}.}

%
%%%%%%%%%%%%%%%%%%%%%%%%%%%%%%%%%%%%%%%%%%
\section{Magnetars}
\label{sec:magnetars}
Soft Gamma Repeaters (SGRs) and Anomalous X-ray Pulsars (AXPs) are thought to be magnetars, that is, young neutron stars (NSs) with extremely high magnetic fields~\cite{Katz82, DuncanThompson92, ThompsonDuncan95} and are among the candidates for the sources of FRBs. About thirty magnetars [\url{http://www.physics.mcgill.ca/~pulsar/magnetar/main.html}] (accessed on 1 March 2021) are currently known in our Galaxy (and the Magellanic Clouds), five of which exhibited transient radio pulsations.
The recent detection of $\gamma$-ray emission simultaneous to a fast radio burst (FRB~20200428A) originated in the Galactic SGR~J1935+2154 has demonstrated the common origin of these phenomena. However the energetic for this event is of the order of $10^{-6}$ times that of a cosmological FRB at $z \sim 1$. We should point out that recently bursts just one decade more energetic than FRB~20200428A were observed for FRB~20180916B~\cite{Marthi20}, so it is not clear if it represents just the tail of a population, as volumetric-rate estimates might suggest~\cite{Lu20}. Assuming this is the case, not only must emission models be able to explain the extremely wide range of radio fluxes, but also the radio-to-$\gamma$-ray fluence ratio of FRB~20200428A ($\simeq 2$--$4\times 10^{-6}$ in~\cite{Chime20SGR1915} and, more reliably, $3\times 10^{-5}$ in~\cite{Bochenek20Nat}), which is more than five orders of magnitude greater than that of SGR~1806$-$20 as no FRB was observed in the giant 27 December 2004 outburst of this SGR~\cite{Tendulkar16}.
The Galactic FRB~20200428A is by far the most radio-luminous such event detected from any Galactic magnetar.
The brightest radio burst previously seen from a magnetar was during the 2009 outburst of 1E 1547.0--408 and was three orders of magnitude fainter. Thus, FRB~20200428A clearly suggests that magnetars can produce far brighter radio bursts than has been previously known.

The prominent role of magnetars as promising candidates for extragalactic FRB sources has fostered a number of complementary attempts to identify counterparts or associations with other classes of known sources: since magnetars are believed to represent the endpoint of some core-collapsed progenitors of long GRBs (e.g.,~\cite{Troja07, Lyons10, DallOsso11}), as well as the result of a compact binary merger signalled by a short GRB (e.g.,~\cite{FanXu06, Rowlinson10, Rowlinson13}), some of these GRB sources were targeted by radio follow-up observations, either within hours of the GRB or years later, to search for FRB emission~\cite{Madison19, Men19, Rowlinson19, Rowlinson20, Bouwhuis20, Palliyaguru21}. Systematic and sensitive searches for emission compatible with MGFs from well localised FRB sources have also been carried out in parallel, both independently of and simultaneously with radio observations, whose results and implications are presented in Section \ref{sec:SGR1935}.

%
%%%%%%%%%%%%%%%%%%%%%%%%%%%%%%%%%%%%%%%%%%
\section{Host Galaxies}
\label{sec:Hosts}
To date the detection of FRBs with associated small (arcsec) error boxes have allowed the detection of thirteen putative host galaxies [\url{http://frbhosts.org/}] (accessed on 1 March 2021) with a luminosity distances range from 149 Mpc to 4 Gpc.
Not only has this given solid bases to their cosmological origin, but has also enabled the possibility to explore the host galaxy population, their global properties and the local FRB environment, which are crucial in understanding FRB progenitor systems.
Additionally, the association of a FRB with an optical/NIR host galaxy allows us to get precise measurements of the redshift as well as indirect, but fundamental, information on the nature on the progenitor systems and on the intervening medium toward the observer.
No association was obtained in the early years because of the arcminute localisation of FRBs due to the use of large single dish telescopes, such as Parkes and Arecibo, while the rapid MWL follow-ups to detect the analogs of GRB afterglows did not produce any reliable counterpart (e.g.,~\cite{Petroff15, Tominaga18}).
The host galaxies of two bursts, FRB~20110214A (${\rm DM} = 168.8$~pc~cm$^{-3}$) and FRB~20171020A (${\rm DM} = 114$~pc~cm$^{-3}$), were extensively searched since their very low dispersion measure (DM) confined the search volumes. The search in archival images and cross matching with several catalogues (e.g., the Vista Hemishere Survey~\cite{McMahon13}; the 2MASS Survey~\cite{Skrutskie06}; the NASA Extragalactic Database [\url{http://ned.ipac.caltech.edu/}] (accessed on 1 March 2021)) singled out plausible candidates, though the large FRB localisation uncertainties did not produce reliable identifications~\cite{Shannon18, Petroff19}.
However, a potential host galaxy association for both FRBs came up from subsequent further archival searches (WISE, DSS2, VISTA, NED, SkyMapper [\url{http://skymapper.anu.edu.au}] (accessed on 1 March 2021)) and dedicated spectroscopic follow-up observations (for FRB~20171020A). The WISE~J0120--4950 galaxy, a late-type star-forming galaxy at an estimated redshift $z \sim 0.1$ is the most convincing putative host of FRB~20110214A~\cite{Lee19}, while the bright Sc galaxy ESO~601--G036 ($M \sim 9 \times 10^8\; M_\odot$, ${\rm SFR} \sim 0.13\; M_\odot$~yr$^{-1}$, $z \simeq 0.0867$ is the most likely host of FRB~20171020A~\cite{Mahony18}.

\begin{figure}[H]

\includegraphics[width=0.95\columnwidth]{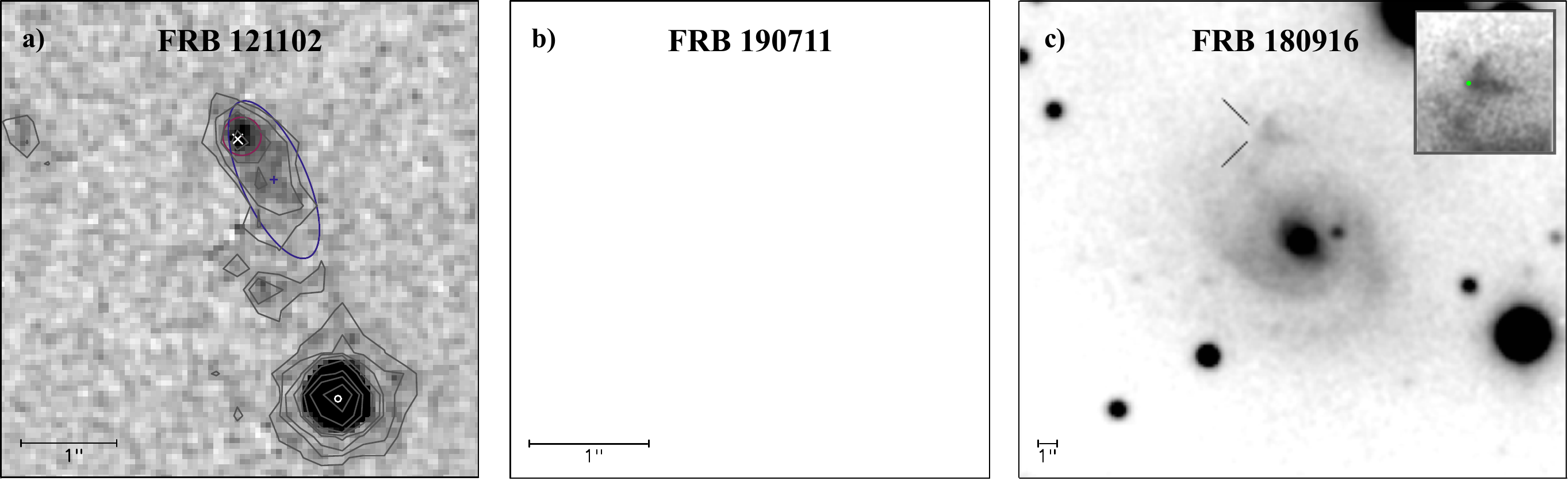}
\caption{\textls[-15]{The three repeaters with an identified host galaxy.
({\bf a}) {\em HST}-IR image of the FRB~20121102A host galaxy (adapted with permission from Bassa, C., et al.; published by IOP Publishing, 2017 \cite{Bassa17}). The white cross mark the FRB~20121102A position. The red circle and the blue ellipse denote the half-light radius of the bright knot and of the extended diffuse emission, respectively. The contours indicate the extent of the host galaxy.
({\bf b}) {\em HST}-IR image of the FRB~20190711A host galaxy (adapted with permission from Mannings, A. G., et al.; arXiv preprint, 2020 \cite{Mannings20}).The ellipse marks the FRB position (2$\sigma$ uncertainty in each coordinate).
({\bf c}) Gemini-North FRB~20180916B host galaxy image ($r^\prime$ filter) taken from the
Public Gemini Observatory Archive (\url{https://archive.gemini.edu}, Program ID GN-2019A-DD-110; see \cite{Marcote20}).
The star-forming region containing the FRB~20180916B position (small green circle) is zoomed-in in the inset.
In all images North is up and East to the left.}
\label{fig:rFRB_hosts}}
\end{figure}

Clearly repeating FRBs offer easier chances for precise localisation by using interferometers. Indeed, the first accurate localisation was that of the repeating FRB~20121102A with the VLA~\cite{Chatterjee17}, which occurred in a low-metallicity, dwarf galaxy ($M = 1.48 \times 10^8\; M_\odot$), projected on a persistent, radio-emitting star-forming region non-coincident with the nucleus (Figure~\ref{fig:rFRB_hosts}{a})~\cite{Bassa17}. The properties of the host galaxy showed remarkable similarities with the host of LGRB and SLSNe, supporting the hypothesis that FRBs are produced by young millisecond magnetars.
FRB 180916 was localised with milliarcsecond accuracy thanks to VLBI observations that recorded four bursts on June 2019~\cite{Marcote20}. The source was localised in a massive nearby spiral galaxy ($M = 2.15 \times 10^9\; M_\odot$, $z = 0.0337$) on a star-forming region with no persistent radio emission (Figure~\ref{fig:rFRB_hosts}{c}). These findings showed that repeating FRBs may originate from diverse host galaxies and local environments.

FRB~20170107A is the first FRB detected by the Australian Square Kilometre Array Pathfinder (ASKAP) and was observed in $i$-band with IMACS on the 6.5 m Magellan Baade Telescope to search for the host galaxy~\cite{Eftekhari18}.
Driven by the properties of the FRB~20121102A persistent radio counterpart, star-forming galaxies were excluded from the search setting a lower cut on the radio-to-optical brightness ratio for the persistent source to $S_{\rm 1.4 GHz}/S_V~=~25$.
In the 5\farcm3 $\times$ 4\farcm2 error region two candidate hosts were identified with a brightness ratio $\gtrsim 100$. We now know that FRB~20121102A represents an exception, so the assumption made in this study need to be revised and eventually the three additional star-forming host galaxy candidates found are worth additional investigations.

FRB localisations have dramatically improved in the last few years thanks to the entry into service of interferometers such as ASKAP/ICS and DSA-10, that achieve (sub)arcsecond positions over field of view (FoV) of several tens of square degrees.
Bannister et al.  (2019)~\cite{Bannister19} reported the discovery of the one-off FRB~20180924B inside a massive ($M \simeq 2.2 \times 10^{10}\; M_\odot$), $r = 20.54$~mag early-type spiral galaxy at $z = 0.3214$ with an estimated SFR upper limit of <$2.0\; M_\odot$~yr$^{-1}$, hence dramatically different from that of FRB~20121102A.
FRB~20190523A was detected by DSA-10 and localised to a few-arcsecond region containing a single galaxy (PSO~J207+72) at a redshift $z = 0.660 \pm 0.002$, compatible with its DM~\cite{Ravi19}. The galaxy parameters were derived by modelling the Pan-STARRS photometry and the KechI-LRSI spectroscopy with a resulting high stellar mass ($M \simeq 1.2 \times 10^{11}\; M_\odot$) and a low SFR ($\lesssim 1.3\; M_\odot$~yr$^{-1}$). A more stringent upper limit of $< 0.09\; M_\odot$~yr$^{-1}$, and stellar mass $M = (61.2 \pm 40.1) \times 10^{9}\; M_\odot$, was later reported~\cite{Heintz20}.

Since then, the Commensal Real-time ASKAP Fast Transients (CRAFT~\cite{Macquart10}) has started to localise routinely both repeating and one-off FRBs to arcsecond accuracy at a frequency of about 5 per year~\cite{Bannister19, Macquart20, Bhandari20hosts, Heintz20}.
All localised FRBs fell within $1^{\prime\prime}$ of an $r < 22$~mag galaxy, for which it has been possible to conduct targeted MWL follow-ups.
Subsequent deep {\em HST} observations strengthened previous FRB-host associations and excluded the presence of satellites or background galaxies~\cite{Mannings20, Chittidi20, Tendulkar21}.

%
%%%%%%%%%%%%%%%%%%%%%%%%%%%%%%%%%%%%%%%%%%
\subsection{The Parent Population}
It is now possible, therefore, to perform systematic studies to investigate the nature of the progenitor systems on (still limited) samples of host galaxies (6 hosts in~\cite{Bhandari20hosts}, 9 in~\cite{LiZhang20}, 13 in~\cite{Heintz20}, 10 in~\cite{Mannings20}, 10 in~\cite{Safarzadeh20}).

\begin{figure}[H]

\includegraphics[width=0.95\columnwidth]{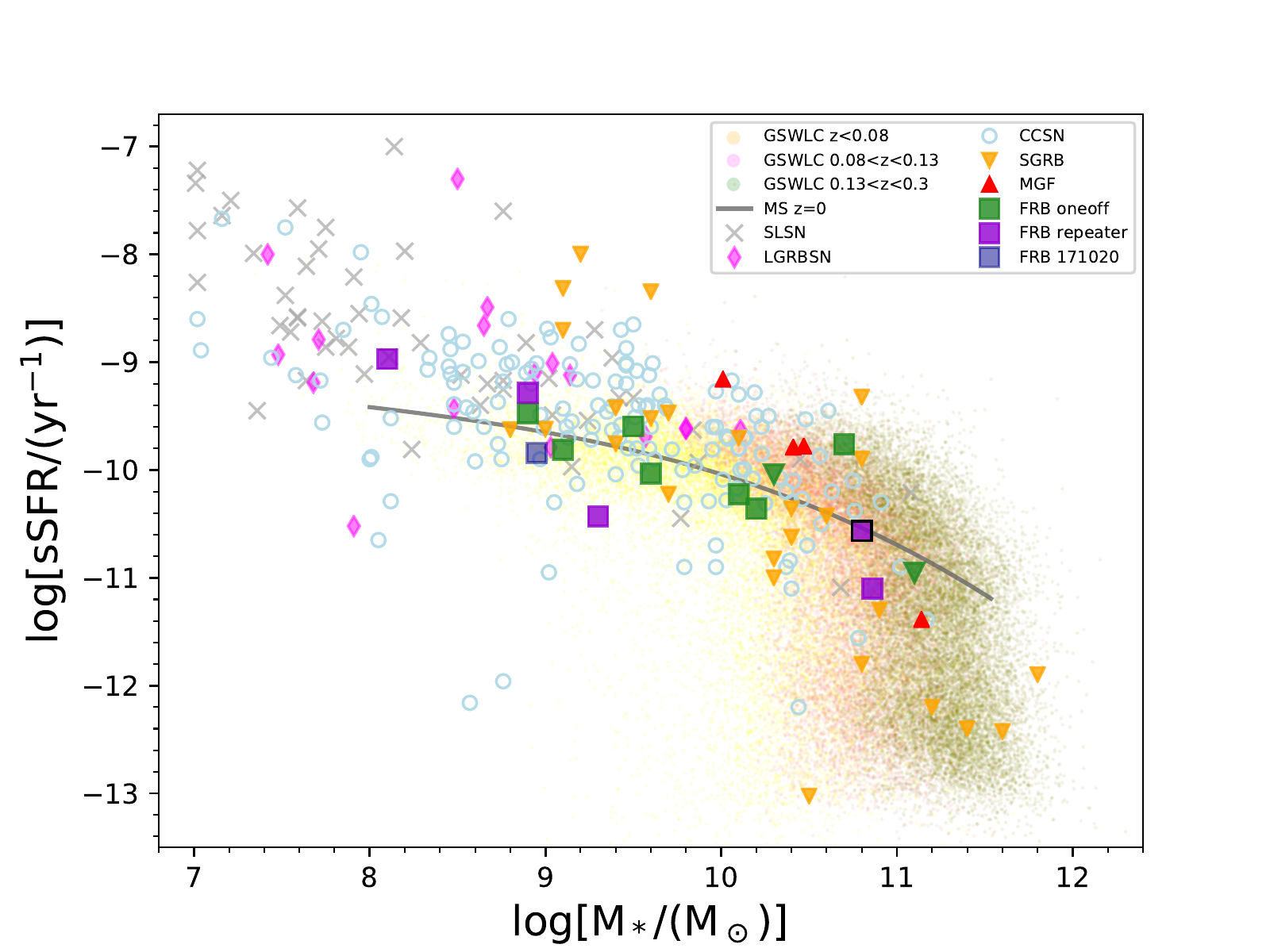}
\caption{\textls[-20]{Specific SFR (SFR/M$_*$) plotted against stellar mass for the FRB hosts and the galaxy populations of other transients. SFR and stellar mass values to derive the specific SFR are from~\cite{Heintz20,Bochenek21mag} (FRB~20180916B),~\cite{Bhardwaj21} (FRB~20200120),~\cite{Bannister19} (FRB~20180924B),~\cite{Ravi19} (FRB~20190523A).
For the two latter FRBs the reported SFR values mark the upper limits (UL).
The FRB hosts are indicated by filled squares (green for the one-off, purple for the repeaters) and filled green triangles for the two UL values. The associated errors are smaller than the size of the symbols. The two hosts with the highest masses among the repeaters are the Milky Way, hosting FRB~20200428A/SGR~J1935+2154 (framed purple square), and M~81, likely hosting FRB~20200120E, respectively. We also added FRB~20171020A among the hosts of the one-off FRBs (blue square) which anyway still is  a putative host.
The galaxies in the GSWLC catalogue at three redshift ranges ($z < 0.08$, $0.08 < z < 0.13$ and $0.13 < z < 0.3$, yellow, pink and dark-green light dots) have been used to represent normal field galaxies.
The grey crosses are SLSNe-I hosts, the filled pink diamonds are LGRBs hosts~\cite{Taggart19}. Short GRB hosts~\cite{Berger14, Fong16}, including the NGC~4993 galaxy (the host of GRB~170817A/GW~170817)~\cite{Pan17}, are shown as orange triangles. The red triangles represent the hosts of magnetar giant flares associated with short GRBs~\cite{Burns21}.
The dark-grey line follows the star formation MS at $ z = 0 $ (as parametrised in equation 5 by~\cite{Saintonge16}).}
\label{fig:hostsM_vs_sSFR}}
\end{figure}

\textls[-15]{Figure~\ref{fig:hostsM_vs_sSFR} shows specific SFRs (sSFRs) plotted against stellar masses of confirmed FRB hosts compared to normal field galaxies from the GALEX-SDSS-WISE Legacy Catalog (GSWLC)~\cite{Salim16} at $z < 0.3$ and to galaxy populations hosting other transients sources~\cite{Taggart19, Berger14, Fong16, Pan17, Burns21} to look for possible progenitors association. SLSNe-I and LGRBs are usually considered as representative of the population of millisecond magnetars in engine-driven SNe, short GRBs as tracers of millisecond magnetars through NS–NS mergers, and normal CCSNe are the dominant formation channels of magnetars in the Milky Way (MW).
We also added a small sample of four local galaxies (namely NGC253, M82, M83 and LMC) hosting a short GRB (SGRB) firmly associated to MGF with radio emission similar to cosmological FRBs~\cite{Burns21}.}

\textls[-15]{As already stated above the hosts of rFRBs show a broad range of galaxy properties thus suggesting possible different progenitor scenarios for FRB events.
The host of FRB 121102 is a low metallicity, dwarf, star-forming galaxy sharing similar properties with the LGRBs and SLSN-I hosts. The FRB~20190711A host is a regular star-forming galaxy (Figure~\ref{fig:rFRB_hosts}{b}) at the high redshift tail of the FRBs sample with the FRB event possibly associated to a CCSN magnetar. Finally the host of FRB~20180916B is a quiescent, massive, spiral galaxy in which the position and characteristic of the FRB location is consistent with the FRB being associated to either a magnetar born in a CCSN~\cite{Bochenek21mag} or a NS in a high mass X-/$\gamma$-ray binary system~\cite{Tendulkar21}.
The inclusion of the MW (and M81) among the rFRB hosts enlarges the range of masses and makes the distribution more similar to that of the one-off FRBs.}

The general picture drawn from Figure~\ref{fig:hostsM_vs_sSFR} is that the FRB hosts are normal galaxies, following the star formation main sequence (MS) at $ z = 0 $~\cite{Saintonge16} and, taking into account also previous results from demographic and statistical studies~\cite{Heintz20, Bochenek21mag, Mannings20, Safarzadeh20}, suggesting a possible consistency with the hosts of CCSNe. However, we remark that the FRB hosts sample is still quite small thus alternative magnetar formation channels or different progenitors may turn up once a larger sample will be available.

Despite the different approaches, all studies converge toward a number of similar~conclusions:
\begin{itemize}
\item FRB hosts span the full continuous range of the main stellar parameters covered by the general sample of galaxies at the same redshifts (typically $z < 0.5$), such as color, stellar mass, SFR~\cite{Heintz20};

\item including the Galactic magnetar among the repeating FRBs there is not a clear differentiation between their hosts and the one-off hosts properties~\cite{LiZhang20};

\item FRB hosts are metal-rich ($12 + {\rm log}(O/H) = 8.7$--9.0) with the noticeable exception of the host of FRB~20121102A (8.08) but globally FRB hosts are consistent with mass-metallicity relation of the field galaxy population at low-$z$~\cite{Heintz20, LiZhang20};

\item the FRB hosts range from starburst to nearly quiescent~\cite{Heintz20}. However, high-resolution imaging shows that most FRBs do not occur in regions of very high SFR compared to the mean values of their hosts~\cite{Mannings20};

\item the majority of FRB hosts show emission lines with a high incidence of LINERS~\cite{Heintz20};

\item FRBs do not occur in the nuclei of the hosts~\cite{Bhandari20hosts, Heintz20};

\item 5 out of 8 host galaxies imaged at high spatial resolution show arm structure and the FRBs are associated to the arms~\cite{Mannings20};

\item the spatial distribution of the FRBs inside their galaxies is not consistent with those of LGRBs and H-poor SLSNe, while better agreement is obtained for CCSNe and SNe Ia~\cite{Bochenek21mag, Heintz20, LiZhang20, Mannings20, Safarzadeh20}. The distribution of SGRBs has a longer tail at large distances from the host centers.
\end{itemize}

Studies of the global properties of the hosts and of the FRB locations inside them strongly disfavour FRB models involving active galactic nuclei (AGNs) and black holes, in general.
FRB progenitor systems do not seem strongly correlated with the most massive stars, thus favouring magnetar models in which the neutron stars are formed via binary neutron star (BNS) mergers, accretion-induced collapse (AIC) of white dwarfs and regular SNe, with respect to those involving prompt, rapidly spinning magnetars~\cite{Heintz20, LiZhang20, Safarzadeh20}.

All previous conclusions are derived from a still small, early sample of FRB hosts.
Precisely-localised FRBs are now detected at a growing rate of several per year. Soon it will be possible to make significant progress toward a stronger link between FRB progenitor systems and their parent populations.
The Milky Way appears as a typical FRB-host galaxy, thus the connection between the recent radio burst from SGR~J1935+2154 and low-luminosity FRBs does not come as a surprise.

\subsection{The Baryon Content of the IGM}
\textls[-15]{The determination of precise FRB distances derived by optical/NIR spectroscopy is a key ingredient to directly probe the baryon content of inter-galactic medium (IGM).
We know that the approximate relation between the DM and $z$, ${\rm DM} \sim 1000 \times z$ pc cm$^{-3}$~\cite{Ioka03, Caleb16, Petroff16, Zhang18_highz_dm} (but see also the more recent estimates in~\cite{Arcus21, Zhang21_distz}), holds once the Galactic component is removed.
In fact, the total observed DM for any FRB can be decomposed in 3 primary components
${\rm DM_{FRB}}(z) = {\rm DM_{MW}} + {\rm DM_{cosmic}}(z) + {\rm DM_{host}}(z)$
where ${\rm DM_{MW}}$, the contribution from our Galaxy, is due to different phases of the gas both in the disk and in the halo, each of the order of $\sim 50$~pc~cm$^{-3}$.
These contributions can be modelled (e.g.,~\cite{Yamasaki20, Cordes02, Yao17}) while Das et al. (2021)~\cite{Das21} have proposed a different approach based on the X-ray absorption.
${\rm DM_{host}}$, the contribution from the host galaxy, includes the host inter-stellar medium and gas local to the FRB. This component, often assumed to be of the same order of magnitude of the Galactic component, can be estimated via measurements of optical emission lines (e.g.,~\cite{Tendulkar17, Chittidi20}).
The contribution from the extragalactic gas is DM$_{\rm cosmic}$(z, $\Omega_b$), where $\Omega_b$ is the cosmic baryon density.
Therefore, pooling together radio and optical observations, it is possible to get estimates of the cosmic baryon density.
The analysis based on the available sample is consistent with values derived from the cosmic microwave background and from the Big Bang nucleosynthesis ($\Omega_b = 0.051_{-0.025}^{+0.021} \times h_{70}^{-1}$~\cite{Macquart20}).
There is, therefore, the promise that soon an enlarged sample of accurately localised FRBs will allow independent measurements of the baryons in the Universe to be performed.}
%

%
%%%%%%%%%%%%%%%%%%%%%%%%%%%%%%%%%%%%%%%%%%
\section{FRBs Multiwavelength Searches}
\label{sec:FRBs_MWLsearches}

Searching for the FRB counterparts at all wavelengths is a crucial task on the road to uncovering their progenitors, emission mechanisms and evolution. The observing strategy can be different depending on the type of phenomenology we want to investigate, though the detection of a fast transient remains the most wanted one.
Archival data, at all wavelengths, have been a valuable resource for several searches and statistical studies: {\em Beppo}SAX, {\em Swift}/BAT, {\em Insight-HXMT}, {\em Fermi}/GBM, and so forth, in the X-ray band and Zwicky Transient Facility (ZTF, previously PTF), {\em Gaia}, Evryscope, and so forth, in the optical band (details are reported below). In fact knowledge of the exact time of an event allows a dedicated refined analysis to be conducted, eventually combining data from different sources, thus allowing for potential sub-threshold detections to be pinpointed.
Whenever a FRB is detected with a relatively small error box (typically sub-arcmin), the attempt to identify uniquely its host galaxy and any possible simultaneous/delayed transient or persistent emission is facilitated as dedicated and/or large telescopes with a small FoV instrumentation can come into play. Still, both wide-area and pointed MWL monitoring campaigns remain crucial, as they address different aspects of the FRB phenomenology, for example, location/association and time evolution. These are crucial to identify the FRB progenitor(s) and the nature of the emission mechanism(s).
For particularly interesting cases, like the two ``golden'' repeaters FRB~20121102A and FRB~20180916B, dedicated monitoring campaigns can be organized with the participation of ground and space instrumentation. The possibility to concentrate the observational efforts in time windows where the bursting activity is particularly high represents a unique opportunity to perform a simultaneous MWL coverage from the MHz--GHz in the radio to the very-high energies of the Cherenkov telescopes. At our knowledge, no such coordinated ultra-wide-band campaign has been performed for one of the two mentioned FRBs.

With the aim to exploit rapidly decaying high-energy transients, related to FRBs and other transients events, new Target of Opportunity (ToO) operational capabilities were implemented by the {\em Swift} satellite team.
The GUANO pipeline~\cite{Tohuvavohu20GUANO} can autonomously recover the BAT event data around the event time and issue an automatic, highest urgency, ToO request to point XRT and UVOT in principle within 14 min or better.
A demonstration of these capabilities was the rapid follow-up of a VLA/{\em realfast} (a system at the VLA) for commensal fast transient searches~\cite{Law18}) FRB candidate for which the fastest X-ray/UV follow-up of any radio transient was achieved: 32 min~\cite{Tohuvavohu20demo}, though no candidate was detected in the XRT and UVOT data, with $3 \sigma$ upper limits of $F < 3.33 \times 10^{-13}$ erg~cm$^2$~s$^{-1}$ (0.3--10 keV) and $u > 22.18$~mag.
Much shorter reaction times can be achieved for example if a FRB trigger is received near in time to an already existing {\em Swift} commanding pass. In~this case observations could begin in as little as 5 min~\cite{Tohuvavohu20GUANO, Tohuvavohu20demo}.

A reverted strategy where a radio telescope shadows the pointings of an X-ray telescope would be an interesting approach too. Of course the relative instruments FoV is to be taken into account. As an example, a radiotelescope of $\approx 30$ m diameter has a FWHM beamwidth of $\approx 27.5^\prime$ when observing at 1.5~GHz ($\theta_{\rm FWHM} \approx 1.2 \lambda / D$, where $\lambda$ is the observing wavelength and $D$ is the telescope diameter) that would match the $\approx 24^\prime$ FoV of {\em Swift}/XRT~\cite{Burrows05}. But a smaller dish, with a larger beamwidth, would still be a valuable instrument to detect relatively bright FRBs, and could be used also to shadow larger FoV X-/$\gamma$-ray detectors.

An interesting example of this ``reverse'' approach is represented by the ASKAP follow-up campaign of GRBs detected by {\em Fermi}/GBM~\cite{Bouwhuis20}. Twenty GRBs, four of which short, were followed up with a typical latency of about 1 min, for a duration of up to 11~h after the burst. The aim of the campaign was to investigate the Ravi and Lasky  (2014) model for short GRBs which predicts a delayed FRB-like emission in the range 10--$10^4$~s after the merger of two NSs.
The ASKAP fly’s-eye configuration allows a large sky coverage, hence the possibility to cover the degrees-size error boxes of {\em Fermi}/GBM detected short GRBs, with respect to those detected by {\em Swift}/BAT-XRT (arcmin--arcsec).
In fact each antenna has 36 beams covering a sky area of 30~deg$^2$. Combining the 6--8 antennas used in the campaign results in an instantaneous FoV between 180 and 240~deg$^2$.
For a putative FRB duration in the range $w = 1.265$--40.48 ms, an upper limit of 26~Jy~ms ($w/$1ms)$^{-1/2}$ was obtained for any radio burst arriving in the time range $\sim$2 min--h after the detection of the GRB.
Given the large model uncertainties on the probability of BNS mergers to result in supramassive stars, collapse time distributions and FRB energetic~\cite{RaviLasky14, Zhang14}, the null result does not allow us to draw any conclusion about the possibility to have a FRB associated to a short GRB.
On the other hand if the radio and X-/$\gamma$-ray coverage had been simultaneous it would have allowed us to investigate emission models which predict a quasi-simultaneous/short delay between the radio and high-energy emission.

Searches for FRBs optical or high-energy counterparts were conducted during standard ``triggered'' follow-up observations (e.g.,~\cite{Petroff15, ShannonRavi17, Bhandari18, Tominaga18}) as well as during simultaneous observations with wide-field telescopes (e.g.,~\cite{DeLaunay16, Martone19, TingayYang19, Guidorzi20a}) or targeting the two ``golden'' repeaters (e.g.,~\cite{Scholz16, Scholz17, Hardy17, MAGIC18, Cruces21, Guidorzi20b}).
Various radio, optical, X-ray, and $\gamma$-ray bands coordinated observations have targeted again the two repeaters and periodic FRB~20121102A~\cite{Scholz16, ZhangZhang17, Chatterjee17, Sun19, Cruces21} and FRB~20180916B~\cite{Casentini20, Panessa20, Pilia20, Scholz20, Tavani20, Zampieri20}.
Currently the best optical upper limit on the millisecond optical emission of a FRB is that reported for FRB~20180916B and was obtained by the fast photometer TNG/SiFAP2~\cite{Pilia20}, which is $5.4 \times 10^{42}$~erg~s$^{-1}$ (see Section \ref{sec:FRB1809}). Unfortunately it is not constrained by a simultaneous detection of a radio burst. Observations with similar instruments on larger telescopes could reduce the upper limit by about one order of magnitude. This could be the case of the 10.4-m Gran Telescopio Canarias (GTC) HiPERCAM~\cite{Dhillon18}.
At higher energies, the relatively weak burst luminosity upper limits in the keV and MeV range of $\sim$$10^{45}$~erg~s$^{-1}$ and of $3\sim 10^{46}$~erg~s$^{-1}$ reported by~\cite{Pilia20} will hopefully be superseded by new more stringent measurements from future MWL campaigns.

We want to stress here that since the mechanism(s) governing the emission in the radio and at higher energy is not yet identified, it is plausible that burst times are not synchronous at the millisecond scale. Actually it is very much possible that the higher energy radiation related to a FRB is due to a different emission process so that a delay and even a long lasting afterglow emission (seconds-minutes) cannot be excluded. Assuming that at least a fraction of the FRBs are almost certainly related to magnetar giant flares (see e.g.,~\cite{Yang20, Yang21_1935comp, Zhang20Nat} and references therein), one can think for example, of concentrating MWL campaigns to SGRs monitoring. This is reasonable, however occurrence statistics do not recommend it as a sole approach. On the other hand the wide range of energetic and environmental conditions that are needed to justify the observed radio emission characteristics, like narrow-band emission, frequency drift, bursts not simultaneous as observed at different frequencies (see e.g.,~\cite{Pleunis21} and references therein), and so forth, imply that searches of simultaneous radio/higher-energies bursts are even harder to be conducted than in the radio band alone. This leads us also to conclude that any future detection will likely remain subject to criticisms, independently from its signal-to-noise ratio statistical significance.

In this section we summarize the main observational efforts at wavelength other than the radio (with some exceptions) and show the main results obtained.
We anticipate that, so far, no counterpart at optical wavelengths was reported, whereas a single, very low significance hard X-ray detection with the {\em Swift}/BAT remains debated~\cite{DeLaunay16}.
%

%
%%%%%%%%%%%%%%%%%%%%%%%%%%%%%%%%%%%%%%%%%%
\subsection{MWL Emission Models}
Yi et al. (2014)~\cite{Yi14} applied to FRBs the standard external shock synchrotron emission afterglow model of GRBs to predict the MWL emission evolution. Adopting a simple standard fireball model with a fixed Lorentz factor $\eta = 100$, a number density of the ambient medium $n_0 = 1$ cm$^{-3}$, and making typical assumptions for the other model parameters, they calculated the afterglow light curves in the X-ray, optical and radio showing that the broad-band FRB afterglows are all very faint except in cases of large energies ($E \gtrsim 10^{47}$ erg) and a small redshifts ($z \lesssim 0.1$). To note that while the forward shock emission component is always present, a bright reverse shock emission component is highly magnetization parameter dependent (see~\cite{Yi14} for details). In all cases, comparison with the sensitivity of {\em Swift}/XRT, the Vera Rubin telescope (LSST) and the Expanded Very Large Array (EVLA) show that the accessible afterglow parameters space is quite small.

The synchrotron maser emission model proposed by Metzger et al. (2019)~\cite{Metzger19} considers a magnetized relativistic shock as a mechanism for FRBs (see also~\cite{Lyubarsky14, Beloborodov17}). The shocks are generated by the deceleration of ultra-relativistic
shells of energy, likely produced by a central compact object, by a dense external environment.
To explain the FRB~20121102A persistent emission and high rotation measure, the external medium is assumed to be a sub-relativistic electron–ion outflow, instead of an ultra-relativistic wind.
Among other things, the model predicts a (incoherent) synchrotron afterglow, but unlike normal GRB afterglows the emission is produced by thermal electrons heated at the shock rather than a power-law non-thermal distribution.
The emission peaks at hard $\gamma$-ray energies on a time-scale comparable or shorter than the FRB itself, with a time-scale of the order of seconds in the X-ray band (see {Figure}~8 in Metzger et al. (2019)~\cite{Metzger19}). %Please confirm whether it is figure 8 in the references? or Modify the order of figure mention Mention figure in order
%R: Made the reference to the paper explicit
The predicted peak luminosity are $L_\gamma \sim 10^{45-46}$ erg~s$^{-1}$ in the MeV--GeV range and $L_{\rm X} \sim 10^{42-43}$ erg s$^{-1}$ in the 1--10 keV X-ray band.
Unfortunately, for flare energies in the range needed to explain
the properties of observed FRBs, this signal is challenging to detect
with current $\gamma$-ray and X-ray satellites, even at the estimated
distances of the closest repeating FRB source.
Prospects could be better in the visual band if the upstream medium of the shock has a much higher density like in the dark phases right after major flares, or if the upstream medium is loaded with a large number of $e^+ / e^-$ pairs (e.g., from a rotationally powered component of the magnetar wind).

As mentioned, the emission models library is still large and we'll mention some more of them whenever appropriate.

%
%%%%%%%%%%%%%%%%%%%%%%%%%%%%%%%%%%%%%%%%%%
%\subsection{FRBs optical/NIR observation
\subsection{Past and Ongoing Searches of Optical/NIR FRB Counterparts}
\label{sec:optical-studies}

Search strategies for the optical/NIR counterpart of FRBs are mostly derived from the experience built on other transient sources, namely GRBs. In fact even if the existing alert systems used by $\gamma$-ray satellites to publicly distribute GRBs error boxes (GCN, VO-event) are not commonly used by the radio community (typically private collaborations are in place), the observational approach aimed at detecting a quasi-simultaneous emission or to search for a delayed or constant emission by a possible FRB counterpart are the same.
While a relevant number of dedicated facilities for the optical/NIR detection and/or follow-up of MWL transients (multi-messenger in the case of gravitational wave events) exist and have proven very effective, new ones are close to completion, namely MeerLICHT~\cite{Paterson19}, Black-GEM~\cite{Groot19}, and Deeper, Wider, Faster (DWF~\cite{AndreoniCooke19, Andreoni20DECam}). While not specifically designed for searches in the sub-second range, their capabilities can definitely be exploited for FRB searches. The DWF programme for example has more than 40 participating facilities, including the Subaru/Hyper Suprime-Cam (HSC).
Moreover statistical constraints of MWL counterparts to FRBs can also be derived using large sky area monitoring programs, that is, without scheduled simultaneous or coordinated radio observation~\cite{Chen20, Andreoni20, Andreoni20DECam, Tingay20}. This ``commensal usage'' can be applied not only to data collected by telescopes targeting astronomical sources, like the Vera Rubin, but also to those data collected for completely different purposes. A brilliant example is the Space Surveillance Telescope [\url{https://www.ll.mit.edu/r-d/projects/space-surveillance-telescope}] (accessed on 1 March 2021). Several other similar wide-field telescopes already exist or are about to be completed. See for example the NEOSTEL [\url{https://en.wikipedia.org/wiki/NEOSTEL}] (accessed on 1 March 2021)~\cite{Cibin16}.

In terms of FRB optical follow-up observations, the deepest to-date are possibly those of FRB~20151230A (DM estimated $z \lesssim 0.8$) performed by Tominaga et al. (2018)~\cite{Tominaga18} with Subaru/HSC at three post-burst epoch (8, 11, 14 days). The $gri-$band observations consisted of multiple 3.5 or 4 min exposures with dithering, and reached the $\sim$$26.5$ mag limit for point sources ($5\sigma$). Of the 13 variable sources, potential counterpart candidates, found in the $15^\prime$ radius error circle, two were excluded for radio observational constraints discrepancy, eight were consistent with optical variability of AGNs, two resulted compatible with Type IIn supernovae. The final candidate is located off-center of an extended source and was proposed to be a rather peculiar (faint peak and fast decline) rapid transient (RT) located at $z \sim 0.2$--0.4. No candidate light curve could be reproduced with the SN Ia template. Remarkably, the photometric redshifts of the host galaxies of 11 candidates resulted consistent with the maximum redshift inferred from the DM of FRB~20151230A. If the actual redshift of FRB~20151230A is in the range $z \sim 0.6$--0.8, sensitivity issues could justify these findings. Moreover, given that the volumetric rate of RTs (4800--8000 events yr$^{-1}$ Gpc$^{-3}$~\cite{Drout14}) and FRBs are broadly consistent, if this candidate is really an RT it may be related to the FRB. As the redshift of the putative host could not be derived because of the contamination of the transient, it would be interesting to perform additional investigations. The same is valid for the two hosts of the Type IIn candidates.

\textls[-20]{FRBs being detected by the CRAFT survey allow us to identify and study their host galaxies at all wavelengths. Multi-epoch observations can be performed to detect potential (slow) transient or variable counterparts sources.
Marnoch et al. (2020)~\cite{Marnoch20} used the ESO VLT to study the hosts of three FRBs localised by CRAFT: FRB~20180924B~\cite{Bannister19}, FRB~20181112A~\cite{Prochaska19} and FRB~20190102C~\cite{Macquart20}, which have not been found to repeat despite extensive follow-up in the same fields~\cite{James20}.
Monte Carlo analysis and sources non-detection led to the conclusion that it is unlikely that every apparently non-repeating FRB is coincident with a Type Ia or Type IIn supernova explosion, or with another type of slow optical transient with a similar light curve.
Deeper imaging or prompt optical follow-up would be helpful to detect/exclude other types of transients (SLSN or kilonovae).}

The sub-arcsecond ASKAP detected FRB~20191001A is located in the outskirts of a $r = 18.41$ mag, highly star-forming spiral ($\sim$$8 M_\odot$~yr$^{-1}$), in a galaxy pair, at redshift $z = 0.2340 \pm 0.0001$~\cite{Bhandari20lim}. The Australia Telescope Compact Array (ATCA) observations at 5.5 and 7.5 GHz did not find a compact persistent radio source co-located with FRB~20191001A above a flux density of $15\; \mu$Jy. Deep optical imaging and spectroscopic observations of the host were performed with the VLT-FORS2 and the Gemini Multi-Object Spectrograph (GMOS), respectively (see Section \ref{sec:Hosts} and~\cite{Heintz20}).

\textls[-20]{FRB~20180417A was detected by the CRAFT survey while targeting the Virgo Cluster~\cite{Agarwal19Virgo}, motivated by a possible enhancement in FRB rates in the direction of rich galaxy clusters~\cite{Fialkov18}.
Its sky location was constrained to an error box of size $7^\prime \times 7^\prime$ and Agarwal et al. (2019)~\cite{Agarwal19Virgo} discuss about the possibility FRB~20180417A be located in the Virgo Cluster.
The sum of all the DM contributing components (Milky Way, intracluster medium, intergalactic medium and the host) were considered insufficient to account for the FRB high DM of 474.8~pc cm$^{-3}$, leading to the conclusion that it is located beyond Virgo.
Follow-up observations in the optical band were performed about a month later using the PROMPT5 telescope located at Cerro Tololo Inter-American Observatory (CTIO). A series of thirty 40-s R-band images were collected and compared to archival images to search for any variable source in the region.
A 1374 s $r$-band image taken in 2013 by the Canada–France–Hawaii Telescope (CFHT) MegaCam and a 180-s R-band VMOS image taken in 2009, both covering the FRB error box, were retrieved from the CFHT and ESO archives. Using image subtraction and artificial stars injection, no variable sources with S/N $> 3$ were detected with an upper limit of $R = 20.1$~mag.}

FRB~20180924B and FRB~20190523A were localized in the outskirt of their hosts. The position and environment of both FRBs were suggesting a consistency with the populations of SGRB produced by BNS merger~\cite{Margalit19, WangFRB180924}. Recently, Gourdji et al. (2020)~\cite{Gourdji20} explored again this possible scenario by testing different models and searching for a contemporaneous SGRB with a sub-threshold search of {\em Fermi}/GBM data. They ruled out the possibility of either FRBs being produced by a compact object merger but rather by a very young remnant pulsar through rotational energy extraction.

No doubt that, once again, monitoring host galaxies, identified or candidate (better if not too far), to try to obtain a simultaneous (first) MWL detection of a FRB remains a fundamental task to identify a FRB counterpart-progenitor.

The two periodic repeaters FRB~20121102A and FRB~20180916B have undoubtedly received by far the highest attention in terms of searches for an optical flash or afterglow associated to a radio burst. The details are reported in the dedicated Section \ref{sec:R1_R3}. Here we just mention the recent Apache Point Observatory (APO) 3.5-m telescope deep optical search by Kilpatrick et al. (2020)~\cite{Kilpatrick20} with a sequence of $gri$ filter images covering a CHIME detected post-burst epoch [+2.2, +1938.1] s.

%
%%%%%%%%%%%%%%%%%%%%%%%%%%%%%%%%%%%%%%%%%%
\subsubsection{Instruments and Observational Strategies}
\label{sec:opt_instr}
The observational strategies to identify FRB counterpart are for several aspects the same as for other fast transients.
As the characteristics of the optical/NIR emission are model dependent, the observational approach cannot be unique and must focus both on a possible quasi-simultaneous emission and a delayed one. It is not the aim of this work to review all the predictions of the various models, but we can broadly split them in ``prompt'' and ``afterglow''.

The above mentioned Yi et al. (2014)~\cite{Yi14}, within the supramassive NS implosion scenario and with the standard fireball model, predict that, if $10^{45}$~erg of total energy is available in the ejecta, an object at $z = 0.1$ would have a forward shock optical light curve that peaks at $\Delta T \sim 10$~s after the the burst, with a ${\rm R}  \sim 27$~mag, and shows a relatively rapid decay.
If a putative reverse shock exists, then the light curve would peak at $\Delta T \sim 0.5$~s (same as the X-ray peak time), with a ${\rm R} \sim 24$~mag, but with a much faster decay when compared to the forward shock (see Figs in~\cite{Yi14}).
FRB-associated inverse Compton scattering processes that can produce optical flashes were also considered by Yang et al. (2019b)~\cite{Yang19optburst}. They find that for a $\sim 1$~ms optical burst a flux density much lower than 0.01~Jy is expected. The same result is obtained in the case the, likely incoherent, optical emission is due to the same mechanism that produces the FRB, in particular the coherent curvature radiation and maser mechanisms (see their {Figure} 3).
Given these figures, even a wide-field and sensitive telescope like the Vera Rubin ($\simeq$$3^\circ \times 3^\circ$, 15--30~s exposure time, limiting magnitude $\sim 24.5$) would not offer ideal observing characteristics. Only energies $E \gtrsim 10^{47}$ erg would produce flashes with a light curve with a peak flux potentially detectable by a medium size telescope. Optical observations covering the FRB event with short (sub-second) exposures look then a more promising choice also in the case of GRB like emission.

We can identify the following three FRB families and the corresponding strategy for the search and study of FRBs in the optical/NIR band:
\begin{enumerate}
\item Newly detected bursts: fast follow-up and MWL archives searches, in particular if the error box is
\begin{itemize}
\item arcsec size $\longrightarrow$ optical/NIR monitoring with medium-size telescopes and then large telescopes to perform spectroscopy of the potential host galaxy;
\item arcmin size $\longrightarrow$ wide-field medium-size telescopes follow-up, potential host galaxies identification and then again spectroscopy of the selected putative hosts to measure their redshift to be compared with the DM derived distance upper limit.
\end{itemize}
\end{enumerate}

This approach mimics the GRB afterglow search strategy, which in the FRB case has, so far, proven not to be effective.
Regarding the first case, we note that error boxes of $5^{\prime\prime}$ or less are needed to have a high confidence identification of an host~\cite{Eftekhari17} associated to a FRB at $z \gtrsim 0.1$.
For the second case instead, low dispersion measure bursts, such as FRB~20171020A with ${\rm DM} \simeq 114$~\cite{Shannon18}, offer the potential for detailed host-galaxy studies by only selecting the few with redshift compatible with the one estimated from the DM~\cite{Mahony18}.
\begin{enumerate}
\item[2.] Repeaters: like for the previous item, but the  monitoring campaign can focus on targeting known/candidate host galaxies. Eventually MWL campaigns including radiotelescopes can be considered in order to detect events happening during the monitoring.

\item[3.] Periodic: MWL campaigns around the expected peak phase with the most sensitive possible instruments with simultaneous epoch coverage.
\end{enumerate}

In addition to dedicated, but time limited, campaigns e.g., of periodic FRBs, radio searches/monitoring complemented by continuous shadowing by ground instruments seems a very promising tool to catch a possible MWL flash. This is for example the case of the MeerLICHT project born from a Dutch--South African--United Kingdom collaboration~\cite{Bloemen16}. Located at the Sutherland station of the South African Astronomical Observatory (SAAO), the MeerLICHT telescope has a 65-cm primary mirror and a 2.7 square degree FoV, nicely matching that of MeerKAT (South Africa’s SKA precursor radio telescope array) interferometer and representing the first continuous simultaneous radio-optical telescope combination. It is equipped with six filters ($u,g,q,r,i,z$) and a 100 megapixel CCD camera~\cite{Bloemen16}.
The simultaneous sky coverage and relatively short time-scale radio-optical correlations allows searches and study of several classes of astrophysical transients, in particular FRBs. This is for example one of the aims of the MeerTRAP project [\url{https://www.meertrap.org/}] (accessed on 1 March 2021)~\cite{Stappers16, Jankowski20}.
The project was made possible by the commensal approach to MeerKAT science. In fact the FRBs search makes use of the data acquired from the radiotelescope while it executes other science observations [\url{htpp://meerlicht.uct.ac.za} (accessed on 1 March 2021), \url{http://thunderkat.uct.ac.za} (accessed on 1 March 2021), \url{http://trapum.org} (accessed on 1 March 2021)].
This same approach is adopted by the CRAFT/ASKAP survey at the other SKA precursor located in Western Australia.

BlackGEM are a set of three telescopes, identical to MeerLICHT, being commissioned at La Silla Observatory (Chile)~\cite{Bloemen16}. In spite their primary goal is detecting and characterizing optical counterparts of gravitational wave events detected by Advanced LIGO and Virgo, they can also be used in other monitoring campaigns, including FRBs. Having more than one telescope observing the same source would increase the significance of a possible weak detection if it happens to be synchronous. The bad news is that the cameras employed in these facilities are not capable to perform exposures shorter than $\sim$1 s. This prevents the exploitation of events with milliseconds duration as their signal could be diluted when exposing too long.

Among the additional facilities and projects actively involved in the FRBs investigation, though some are general-purpose sky survey instruments, we cite here the following:
\begin{itemize}
\item \textls[-15]{Fast and Fortunate for FRB Follow-up (F4) is an international collaboration endeavored to study host galaxies at all non-radio-bands through dedicated photometric and spectroscopic follow-up observations of all arcsecond localized FRBs [\url{https://sites.google.com/ucolick.org/f-4} (accessed on 1 March 2021), \url{https://github.com/FRBs/FRB}] (accessed on 1 March 2021).}
\item Deeper, Wider, Faster~\cite{AndreoniCooke19, Andreoni20DECam}.
\item ZTF surveys the sky in search for transient sources on a regular basis. The potential usage for FRBs searches was discussed by~\cite{Andreoni20}.
\item The ultra-wide multiple telescopes {\em Evryscope}~\cite{Law15}, {\em Pi of the Sky}~\cite{Mankiewicz14, Cwiek14}, {\em MMT}~\cite{Biryukov15}.
\item Various survey archives are also of interest for transient searches and are routinely investigated for FRB studies. Among them: TESS, Pan-STARRS, SDSS, SkyMapper, ASAS-SNTF, DSS, VISTA, WISE (see e.g.,~\cite{Petroff19, Lee19, Mahony18, TingayYang19, Heintz20}.
\end{itemize}

Arc-second localization of an FRB makes not only possible to study the potential host galaxy, but it also allows us to perform classical point-spread function (PSF) photometry at the source position on standard accumulated images. This in turn allows the telescope to be pushed at its detection limit and, as mentioned above, if multiple telescopes detect a source simultaneously, then even a low significance measurement can become relevant.

However, given their fast transient nature, the usage of fast photometers represents the perfect tool to try to identify sporadic flashes or to search for periodicity, like in the case of pulsars. Consequently all the instruments capable of acquiring frames at a rate in the range 10–1000 Hz are to be considered ideal for FRB searches. Here we give a brief overview of those with demonstrated or potential capabilities taking into account that other instruments with similar characteristics exist but we are not aware of their use related to FRBs.

{At present, three fast optical photon counters are regularly in operation and have already been used for simultaneous MWL campaigns targeting the periodic FRB 180916~\cite{Pilia20}: SiFAP2~\cite{Ghedina18} mounted at 3.58-m Telescopio Nazionale Galileo (TNG) in La Palma, Aqueye+~\cite{Naletto13, Zampieri15} mounted at the 1.82-m Copernicus telescope in Asiago, and IFI+Iqueye~\cite{Naletto09, Zampieri19} fiber fed at the 1.2-m Galileo telescope in Asiago. Other optical instruments based on photon counting detectors were in operation till few years ago (OPTIMA~\cite{Kanbach08}, GASP~\cite{Collins09}, BVIT~\cite{McPhate12}, ARCONS~\cite{Mazin13}) but are no longer available or are not frequently mounted on telescopes at present. SiFAP2, Aqueye+ and Iqueye couple fast single photon Silicon detectors having resolution at or below the nanosecond with a timing system capable of very high absolute time accuracy with respect to UTC ($60 \;\mu${s} for SiFAP2,~\cite{Papitto19}; 0.5 nanoseconds for Aqueye+ and Iqueye~\cite{Naletto09}). The narrow FoV and the possibility to sample the photon stream at or below the millisecond make them particularly well suited to perform pointed searches for short duration optical flashes. SiFAP2 and Aqueye+ were used to detect millisecond pulsations from PSR~J1023+0038~\cite{Ambrosino17}.}

Fast (from 1 to tens of millisecond) photometric observations of FRBs can be performed also with high-speed optical cameras based on CCDs with windowing and fast readout, electron-multiplying (EM) CCDs, or complementary metal-oxide-semiconductor (CMOS) technology. ULTRASPEC, a purpose-built EMCCD camera for high-speed imaging~\cite{Tulloch11}, is mounted at the 2.4-m Thai National Telescope and was operated in its fastest windowed ``drift'' mode (70.7 ms frame duration) to perform a simultaneous radio and optical fast photometric campaign of FRB 121102~\cite{Hardy17}. HiPERCAM, mounted at the 10.4-m GTC, is based on ULTRACAM~\cite{Dhillon07} but offers a significant advance in performance, with 4 dichroic beamsplitters to record $u,g,r,i,z$ (300–-1000 nm) images simultaneously on five CCD cameras~\cite{Dhillon18}. The CCD detectors can reach a frame rate of 1000 Hz in a windowed and binned mode. HiPERCAM was recently used for a simultaneous optical/X-/$\gamma$-ray campaign targeting again FRB 121102, with the shadowing of radio telescopes~\cite{Cruces21}. It was also employed to perform deep and accurate fast photometric observations of other types of sources~\cite{Zampieri19, Rebassa-Mansergas19, Parsons20}.

\textls[-15]{{Other instruments for performing high speed imaging are potentially available, such as AstraLux based on a EMCCD detector~\cite{Hormuth08}, PlanetCam based on a scientific CMOS sensor~\cite{Mendikoa16}, both mounted at the 2.2-m telescope at the Calar Alto Observatory, LuckyCam (CCD camera~\cite{Law06}) mounted at the Nordic Optical Telescope in La Palma, and Wide FastCam (EMCCD camera~\cite{Velasco16}) mounted at the Telescopio Carlos S\'anchez in Tenerife. They were all designed for performing lucky imaging and reaching the diffraction limit of the telescope, but can in principle be operated in windowed mode up to several $\times 100$~Hz to carry out fast photometry. Other high speed cameras in operation at present are SALTICAM, mounted at the SALT telescope and equipped with a fast CCD~\cite{ODonoghue06}, and OPTICAM at the Observatorio Astronómico Nacional San Pédro Mártir, with a scientific CMOS detector~\cite{Castro19}.}}

Clearly, photon counting instruments can reach the highest sensitivity to short duration events but have a narrow FoV. They are then suited for optical campaigns of well localized targets with simultaneous radio coverage. The ``data structure'' is the same as for higher energies (e.g., X-rays) instruments. Single photons are detected and their arrival times are saved in event lists that are then binned to produce light curves for the following analysis. The time bin is limited only by the time accuracy of the instrument (for Aqueye+ and Iqueye it can be as low as 1~ns). Since the expected magnitude of a potential fast optical burst (FOB) decreases with decreasing readout time (see below), the observing strategy is sampling the light curve with a time resolution comparable to the duration of the burst while, at the same time, preserving an adequate counting statistics per bin. Assuming that FOBs associated to FRBs have a comparable duration, a time bin of $\sim$1~ms is appropriate and easily achievable. Optical flashes can thus be detected in bins with counting statistics in excess of the expected sky background Poissonian level, with significance thresholds properly set taking into account the number of trials (bins) of an observation. Foreground events (e.g., cosmic rays, artificial satellites, meteors) may be detected and contaminate the observations. Considering the small FoV of the instrument, to properly filter them out simultaneous observations with more than one instrument and/or at two different sites would be ideal.

\textls[-15]{On the other hand, high speed cameras have typically more limited sensitivity to short duration events. However, their larger FoV makes them the only possibility for monitoring campaigns of less well localized sources or candidate host galaxies. The observing strategy is more similar to that adopted in conventional photometry, but with some important peculiarities (see e.g.,~\cite{Hardy17}). To extract photons from a region that most of the time contains no source, aperture photometry with a fixed-sized aperture is preferable over PSF photometry. The aperture should be sufficiently large to accommodate the positional uncertainty and the average seeing throughout the observation. The FRB position can be determined performing an offset from the measured position of a comparison star in each frame.}

As an example of the different performances with varying instrument, we consider a putative detection of an FOB close in time to an FRB, with a fluence of 5--10~mJy and a duration of 1~ms. The expected FOB magnitude is (see e.g.,~\cite{Lyutikov16, Yang19optburst}):
\begin{equation}
{\rm V} = 16.4 - 2.5 \log ( \tau_{\rm ms}\, F_{\rm mJy} / T_{\rm ms} ) \, ,
\end{equation}
where $F_{\rm mJy}$ and $\tau_{\rm ms}$ are the FOB flux density (in mJy) and duration (in ms), and $T_{\rm ms}$ is the integration or sampling time (in ms). For photon counting instruments with a sampling time $T_{\rm ms} = 1$, and assuming $F_{\rm mJy} = 5$--10 and $\tau_{\rm ms} = 1$, the instantaneous magnitude is: ${\rm V} = 13.9$--14.7 in 1~ms. The detection threshold of the instrument and the corresponding detection significance depend on the number of 1 ms intervals that have been sampled (trials) and hence on the duration of the observation. However, assuming that an FOB is detected close in time to an FRB, the search can be limited to a few seconds (e.g., 10 s) around the time of arrival of the radio burst.

For photon counting instruments the deepest upper limits to the optical fluence to date have been obtained with SiFAP2@TNG (${\rm V} \simeq  15.5$ in 1~ms, 2~mJy~ms) and Aqueye+@Copernicus (${\rm V} \simeq 13.7$ in 1~ms, 12 mJy ms) within the framework of a MWL campaign on FRB~20180916B~\cite{Pilia20}. These limits take into account the observation duration (20~m for SiFAP2, 1~hour for Aqueye+). For SiFAP2, installed on a 4-m class telescope, the 20 m upper limit is sufficiently deep to show that an FOB with the properties assumed here can be detected even not correcting for the number of trials in 10 s. However, even a 2-m class telescope has the capability to detect such an FOB. Rescaling the upper limit of Aqueye+ to the number of trials in 10~s, the limiting magnitude becomes ${\rm V} \simeq 14.35$ in 1ms and the detection of an FOB of the type considered here would then be possible.

\textls[-15]{For a high speed camera with a frame rate of 100 Hz (integration time $T_{\rm ms} = 10$~ms), the same FOB ($F_{\rm mJy} = 5$--10, $\tau_{\rm ms} = 1$) would give an expected magnitude ${\rm V} = 16.4$--17.2 in 10~ms. If the camera has an overall efficiency similar to that of ULTRASPEC and is mounted on a 2-m class telescope as TNT, the reported limiting fluence is $\sim$4.6~mJy 10~ms~\cite{Hardy17}, leading to a limiting magnitude ${\rm V} \simeq 14.7$ in 10~ms. For the search carried out with ULTRASPEC the limit was derived considering only the frames nearest to the radio burst simultaneously detected with the Effelsberg telescope. Clearly, the limit would increase in brightness ($V < 14.7$) if the search is extended to more images around the time of arrival of a radio burst. Therefore, an FOB with the properties assumed here would not be detected. In order to detect it, either the camera has a very fast readout mode ($\sim$$1$~ms) or a 4-m class (or larger) telescope is needed.}

On the other hand, the availability of low-cost high speed cameras based on different technologies and the possibility to mount them on small commercial telescopes opens up the possibility to perform high cadence observations of several FRB sites and obtain interesting limits to their optical fluence.
A $20^\prime \times 20^\prime$ FoV, $1024 \times 1024$ EMCCD camera capable of a frame rate of 50--100 Hz mounted on a 50-cm telescope at the Ondřejov observatory (Czech Republic) reached a limiting fluence of $\sim$10--20~mJy in $\simeq$10--20~ms~\cite{Karpov19}. This corresponds to ${\rm V} = 13.9$~mag in 10~ms, considering the number of frames/trials. A similar low-cost setup and observing approach could be quite easily adopted on other existing optical telescopes for performing monitoring of selected FRBs with arcmin localization and/or for shadowing observations performed at other wavelengths.

%
%%%%%%%%%%%%%%%%%%%%%%%%%%%%%%%%%%%%%%%%%%
\subsection{FRBs X-/$\gamma$-ray Observations and Studies}
\label{sec:X-gamma-studies}
As soon as the extragalactic nature of FRB sources was gradually established and even before the discovery of FRB repeaters, a number of theoretical models suggested possible links with sources of other transient hard X-/$\gamma$-ray events, such as GRBs: the fact that a millisecond newborn magnetar could form in a GRB, which can be either long~\cite{Usov92, ThompsonDuncan95, Bucciantini07, Metzger11} or short~\cite{FanXu06, Metzger08}, makes it a potential candidate for FRB sources. Radio and high-energy emissions could be either simultaneous~\cite{UsovKatz00, Totani13, Wang16} or with some delay either with the FRB preceding the GRB~\cite{PshirkovPostnov10}, or the other way around. In the latter case, the GRB would signal the formation of a supramassive NS, while the delay between GRB and FRB would correspond to the time it takes for the supramassive NS to finally collapse~\cite{FalckeRezzolla14, RaviLasky14, Zhang14}. Regardless of the formation channel, extragalactic magnetars soon appeared to be among the most promising FRB candidates~\cite{PopovPostnov13, Lyubarsky14, Beloborodov17, LyutikovPopov20}, as was finally corroborated by the detection of FRB~20200428A from the Galactic magnetar SGR~J1935+2154~\cite{Chime20SGR1915, Bochenek20Nat}. The fact that they are well known sources of sporadic X-ray bursts and, more rarely, of hard X-/soft $\gamma$-ray giant flares, also triggered searches for extragalactic magnetar high-energy flaring emission associated with FRBs. These searches were carried out through a number of different approaches, which we summarise in the following sections.

%
%%%%%%%%%%%%%%%%%%%%%%%%%%%%%%%%%%%%%%%%%%
\subsubsection{Searches for Prompt X-/$\gamma$-ray Counterparts}
\textls[-15]{A number of independent searches for prompt hard X-/$\gamma$-ray counterparts to FRBs was carried out using data of different past and presently operational detectors and in different energy bands.
Tendulkar et al. (2016)~\cite{Tendulkar16} carried out one of the first systematic searches of this kind for a sample of 15 FRBs that were promptly visible by {\em Fermi}/GBM, {\em Swift}/BAT or {\em Konus}/WIND, ending up with lower limits on the ratio of radio-to-$\gamma$-ray fluence $F_{\rm 1.4\; GHz}/F_\gamma = \eta_{\rm FRB} \gtrsim 10^{7-9}$~Jy~ms~erg$^{-1}$~cm$^{2}$. Moreover, the absence of any FRB-like emission associated with the hard X/soft $\gamma$-ray giant flare of SGR~1806$-$20 turned out to be mostly incompatible with the limits obtained for the FRB sample. As the size of FRB samples began increasing, the first FRB repeater, FRB~20121102A, was discovered~\cite{Spitler16}. Not only did this pave the way to MWL campaigns, which hence provided the first deep X-ray limits~\cite{Scholz16} to both transient and persistent high-energy emission (see Section~\ref{sec:Xpersistent}), but it also enabled its host galaxy identification and consequent determination of its cosmological distance.}

An initial claim for a {\em Swift}/BAT detected $\gamma$-ray transient positionally and temporally associated with FRB~20131104A with $\sim$$3 \sigma$ confidence~\cite{DeLaunay16} later found no confirmation from other observations~\cite{ShannonRavi17}.
Meanwhile, a number of independent, systematic searches for simultaneous X-/$\gamma$-ray emission of FRBs over a range of timescales were carried out by different groups, exploiting different data. Cunningham et al. (2019)~\cite{Cunningham19} analysed data from {\em Fermi}/GBM and LAT and also {\em Swift}/BAT data available for 23 FRBs and constrained the radio-to-$\gamma$-ray fluence ratios over timescales in the range $0.1$ to $100$~s and in different energy bands, from soft to hard (MeV) $\gamma$-rays. Their lower limits on the distance of potentially associated MGFs turned out to be compatible with the constraints inferred from the corresponding DM values.

\textls[-20]{No high-energy counterpart was found for a sample of 41 FRBs promptly visible with the Cadmium Zing Telluride Imager aboard {\em AstroSat} operating in the 20--200~keV energy band on timescales in the range from 10~ms to 1~s, with consequent upper limits to $\gamma$-ray-to-radio fluence ratio that are comparable with the ones previously obtained and described above~\cite{Anumarlapudi20}. In the case of one of the brightest FRB yet measured, FRB~20010724A, also known as ``the Lorimer burst''~\cite{lorimer07}, a devoted and sensitive search carried out with one of the most sensitive GRB experiments at the time, the {\em Beppo}SAX/GRBM (1996--2002), provided stringent upper limits on the possible associated GRB as a function of distance~\cite{Guidorzi19}---the combination of relatively low DM and large fluence suggests it to be a relatively nearby event~\cite{Shannon18}.
A search for prompt $\gamma$-ray counterparts with {\em Fermi}/GBM data over a broad range of timescales (1 to 200~s) was carried out by Martone et al. (2019)~\cite{Martone19}, who modelled the variable background in the various energy bands through a machine learning approach. In addition, they summed the interpolated {\em Fermi}/GBM light curves by aligning them with the FRB time, thus constraining the systematic presence of an associated $\gamma$-ray signal.}

\begin{figure}[H]

\includegraphics[width=0.8\columnwidth]{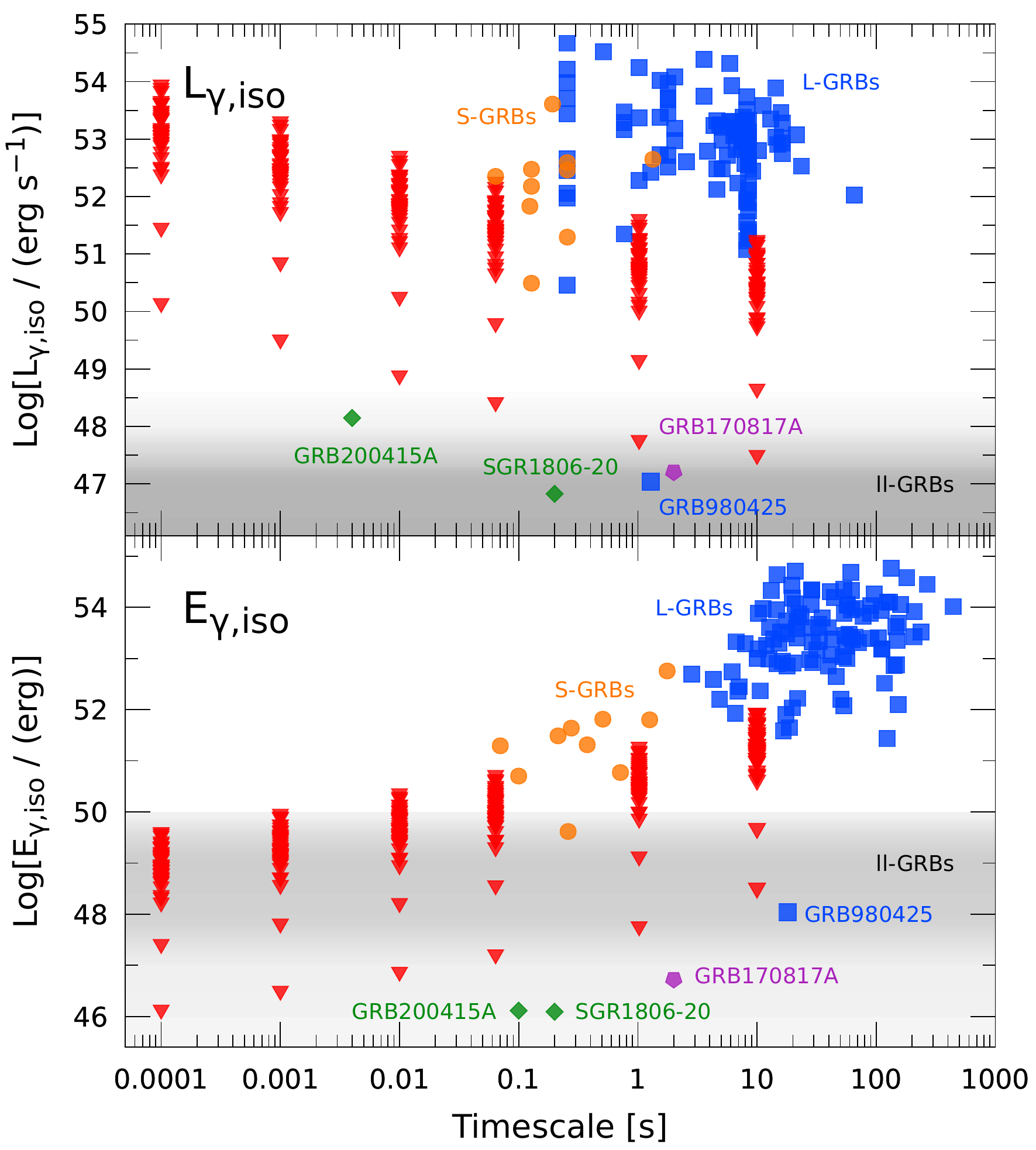}
\caption{Upper limits (red downward triangles) on both isotropic-equivalent $\gamma$-ray luminosity (\textbf{top}) and released energy (\textbf{bottom}) of prompt counterparts to Fast Radio Bursts (FRBs) as a function of timescale, compared with populations of both short (orange circles) and long (blue squares) Gamma Ray Bursts (GRBs). Also shown are the giant flare from the Galactic magnetar SGR~1806--20, GRB~200415A (the magnetar in NGC~253), the short GRB~170817 associated with the first BNS merger detected with gravitational interferometers, and prototypical low-luminosity GRB980425. In the luminosity plot the GRB~200415A spike duration is assumed 4 ms, whereas the $T_{90}$ is 0.1~s~\cite{Svinkin21}. The shaded area shows where most low-luminosity GRBs lie (Figure adapted from~\cite{Guidorzi20a}).
%please confirm if change hyphen as minu sign in the figure
%R: change done.
\label{fig:UL_guidorzi20}}
\end{figure}

The combination of large effective area and exquisite time resolution of the High-Energy instrument aboard {\em Insight-HXMT} enabled an analogous search for prompt simultaneous high-energy counterparts to 39 FRBs in two energy bands, either 40--600 or 200--3000~keV, depending on the operation mode in use at the time of each FRB~\cite{Guidorzi20a}. The explored timescales range from $100~\mu$s to 10~s and ended up with constraining upper limits on the radio-to-$\gamma$-ray fluence ratio. Moreover, in addition to the redshift $z$ information available for three FRBs included in the sample, they exploited the constraints on $z$ derived from DM to obtain upper limits on both luminosity and released energy as a function of timescales. The comparison with typical cosmological short and long GRBs excluded any systematic, simultaneous association with FRBs (Figure~\ref{fig:UL_guidorzi20}).

\begin{figure}[H]

\includegraphics[width=0.8\columnwidth]{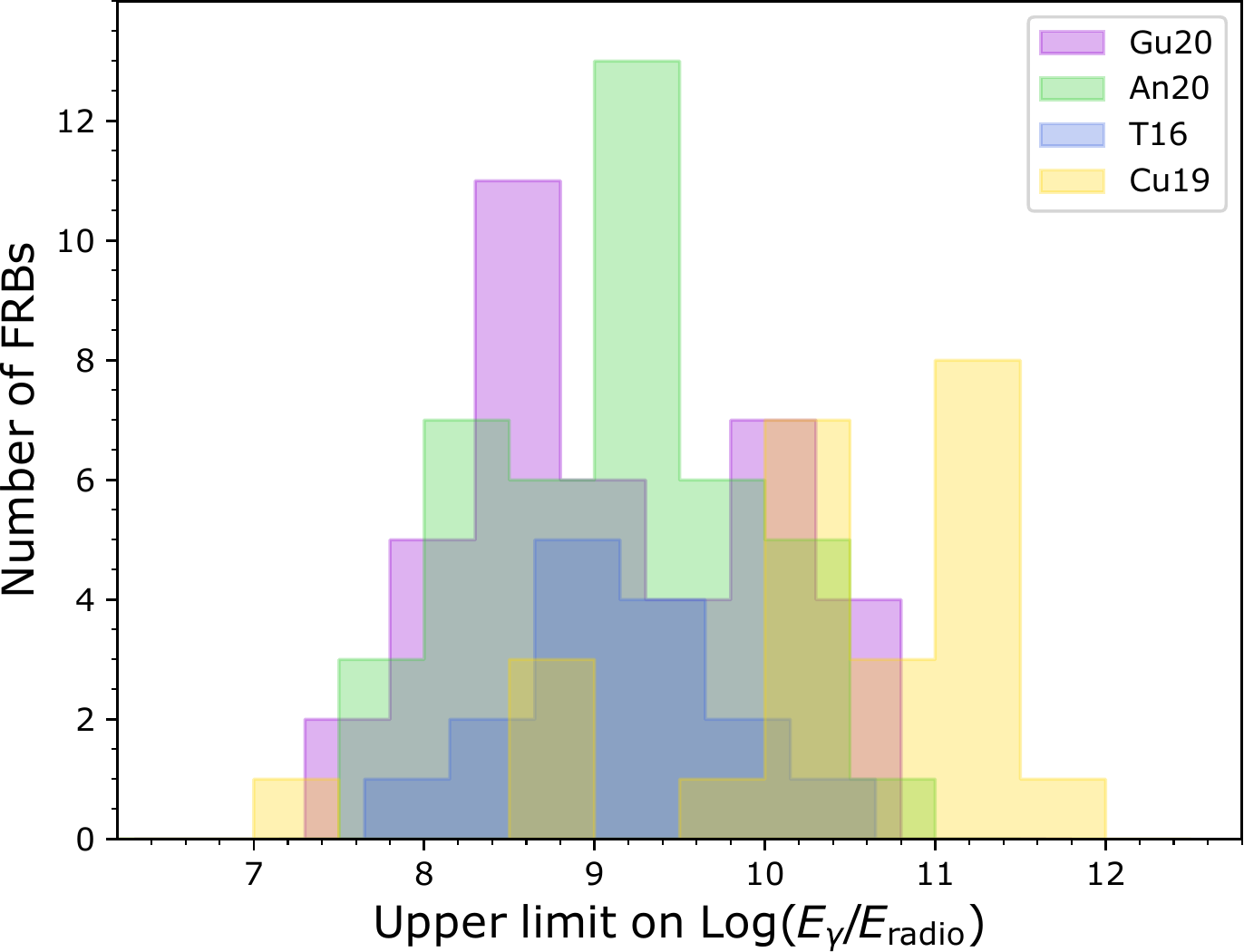}
\caption{Distribution of upper limits on the logarithm of the ratio $E_{\gamma}/E_{\rm radio}$ obtained with hard X-/$\gamma$-ray detectors that were observing at the times of some FRBs for different samples and in different energy bands as reported in the literature: Gu20 refers to {\em Insight-HXMT} data obtained in two different bands, either 40--600 or 200--3000 keV~\cite{Guidorzi20a}; An20 refers to {\em AstroSat} data in the 20--200~keV band~\cite{Anumarlapudi20}; T16 is based on the data from different detectors~\cite{Tendulkar16}; Cu19 is based on {\em Fermi}/GBM data in the 8--$4\times10^4$~keV band, except for the {\em Swift}/BAT data of two bursts in the 15--350~keV band~\cite{Cunningham19}. For all of them an integration time of $0.1$~s was used, except for Gu20 for which 64~ms was used.
\label{fig:Egamma_over_radio}}
\end{figure}

Figure~\ref{fig:Egamma_over_radio} shows the distribution of upper limits to the ratio of $\gamma$-ray-to-radio fluence in case of simultaneous emission, as obtained over different FRB samples with different instruments. We chose the common timescale of $0.1$~s, except for the {\em Insight-HXMT} data, whose closest value is 64~ms. Despite the different energy ranges, data sets, FRB samples, one infers that $E_{\gamma}/E_{\rm radio}< 10^{7-10}$.

The lack of keV--MeV detection of a prompt counterpart to FRBs with measured distance in some cases was significant enough to rule out the possibility that, in some models, FRBs are emitted during the inspiral stage of compact binary mergers involving at least one NS~\cite{Gourdji20}.

Moving to higher energies, in the MeV--GeV range, analogous searches for both simultaneous and subsequent emission have also been carried out. A systematic search within the {\em Fermi}/LAT data for a number of FRBs that went off in the instrument's FOV within a few ms-timescale led to no detection, with upper limits to the ratio $(\nu \, L_\nu)_\gamma/(\nu \, L_\nu)_{\rm radio} \lesssim (4$--$12) \times 10^7$~\cite{Yamasaki16}.
A similar investigation was carried out by
Xi et al. (2017)~\cite{Xi17}, which reported on the search for GeV counterparts to 14 non-repeating FRBs with the {\em Fermi}/LAT, including the mentioned FRB~20131104A.
They find 0.1--100 GeV upper limits in the range of $(0.4$--$190.8) \times 10^{53}$ erg for the isotropic kinetic energy of the possible GRB-like blast wave. To note that this energetic may decrease if the contribution to the DM of FRBs by their local environment and host galaxy contribute significantly, so to decrease the value of the luminosity distance.

In the MeV--GeV band AGILE observed two repeating sources, FRB~20180916B and FRB~20181030A, both promptly and over time intervals as long as 100~days, ending up with upper limits on fluence as a function of the integration times: for example, in the 0.4--100 MeV range the upper limit for FRB~20180916B goes from $10^{-8}$ to several $\times 10^{-7}$~erg~cm$^{-2}$ for integration times spanning from sub-ms to 10~s. A constraint was derived on the released energy on a ms-timescale of $E_{\rm MeV} < 2\times 10^{46}$~erg~\cite{Casentini20}.

While the search for a systematic association of FRBs with GRB sources has so far turned out to be unsuccessful (see e.g.,~\cite{Palaniswamy14, Kaplan15, Bouwhuis20, Palliyaguru21}, an interesting case is offered by FRB~20171209A, which is positionally compatible with a long GRB at $z=0.82$ that was observed with {\em Swift} six years before, GRB~110715A, and whose X-ray afterglow is suggestive of a millisecond magnetar formed in the aftermath of the GRB. Nevertheless, the relatively low statistical significance of the association (2.5--2.6 $\sigma$) makes it somehow questionable~\cite{Wang20}, leaving the possibility of a fake association, as was probably the case for FRB~20131104A and the hard X-ray transient~\cite{DeLaunay16}.

%
%%%%%%%%%%%%%%%%%%%%%%%%%%%%%%%%%%%%%%%%%%
\subsubsection{Constraints on X-/$\gamma$-ray Either Persistent or Long-lived Transient Sources}
\label{sec:Xpersistent}
Concerning follow-up MWL campaigns, one of the first attempts was made for FRB~20140514A: in addition to the numerous radio and optical facilities, the X-ray band was covered with {\em Swift}/XRT, which began observing $8.5$~h after the FRB and found no source down to $8.2 \times 10^{-15}$~erg~cm$^{-2}$~s$^{-1}$ in the $0.3$--$10$ keV energy band. As a result, the presence of an associated typical GRB afterglow was ruled out~\cite{Petroff15}.

The opportunity to look for and constrain either persistent or long-lived transient sources, such as GRB afterglows, greatly benefited from the discovery of repeating sources.
Following the discovery of the first repeaters and sub-arcsec localisation of some FRB sources, the possibility of systematic MWL campaigns blossomed. In the following sections we describe in more detail the campaigns that were devoted in particular to the two most studied repeaters with measured distance, FRB~20121102A and FRB~20180916B. For example, in the case of FRB~20121102A, a deep search for a persistent X-ray source using {\em Chandra} gave a $5 \sigma$ upper limit for the $0.5$--$10$~keV isotropic luminosity of $3 \times 10^{41}$~erg~s$^{-1}$ at the distance of 121102~\cite{Scholz17} (972~Mpc~\cite{Tendulkar17}, see also~\cite{Chatterjee17}).
{\em Fermi}/LAT data taken over eight years were used by to derive an upper limit in the 100~MeV--10~GeV isotropic luminosity of $4 \times 10^{44}$~erg~s$^{-1}$~\cite{ZhangZhang17}.
These upper limits are not tight enough to pose significant constraints on the parameters of emission models either invoking or not a young magnetar as the source of the repeating FRBs.

Concerning the other most studied and much closer (luminosity distance of 149~Mpc) repeater FRB~20180916B, {\em Swift}/XRT observations constrained the 0.3--10~keV cumulative luminosity to $L_{\rm X} \lesssim 1.5 \times 10^{41}$~erg~s$^{-1}$~\cite{Tavani20}.
Sensitive {\em Chandra} observations constrained a possible persistent X-ray source in the band $0.5$--$10$~keV down to a luminosity limit of $L_{\rm X} \lesssim 2 \times 10^{40}$~erg~s$^{-1}$~\cite{Scholz20}.
In the $\gamma$-rays a set of AGILE observations over a many-year timescale yield an upper limit of $L{\gamma} \lesssim 2 \times 10^{42}$~erg~s$^{-1}$ at energies above 100~MeV~\cite{Tavani20}.

In the energy range 0.1--10~GeV, a ten-year upper limit of $7.3 \times 10^{43}$~erg~s$^{-1}$ on the luminosity of a persistent source potentially associated with the repeater FRB~20180814A (FRB~180814.J0422+73) was obtained from {\em Fermi}/LAT data. This poses constraints in the magnetic field-initial spin period of a hypothetical magnetar as well as on the possible emission of a high-energy GRB afterglow~\cite{Yang19}.

%
%%%%%%%%%%%%%%%%%%%%%%%%%%%%%%%%%%%%%%%%%%
\subsection{VHE $\gamma$-rays Observations and Neutrino Events Searches}
The current theoretical and observational investigations of the possible emission of VHE $\gamma$-rays from FRBs is very limited, still VHE observations are useful to constrain present and future emission models. A model based on SGRs was proposed by Lyubarsky et al. (2014)~\cite{Lyubarsky14} and predicted that millisecond VHE emission could be visible at distances up to about 100 Mpc.
As discussed above, FRBs, from the galactic FRB~20200428A to the extragalactic population, are possibly all originated from the flaring activity of magnetars~\cite{Lu20, Margalit20uni}. Among others, a number of variants on the synchrotron maser model were proposed that differ regarding the nature of the upstream medium and the required shock properties, but all predict that FRBs are accompanied by hard radiation counterparts, though with different characteristics.
For example Margalit et al. (2020a)~\cite{Margalit20uni} predict a peak energy range from $\sim$100~keV (Galactic bursts) to $\sim$100~GeV, depending on the flare energy. A relative delay comparable or shorter than the radio
burst duration (i.e., $\lesssim$1\,ms) is predicted between the high-energy burst observed at peak and the radio burst.
To note also that VHE emission, travelling through the extragalactic radiation fields, is attenuated by interactions with the extragalactic background light (EBL) via e$^+$/e$^-$ pair-creation processes. This leads to the collective effect of an absorption of $\gamma$-rays at the highest energies (see e.g.,~\cite{Gilmore12}).

In addition to VHE emission, imaging atmospheric Cherenkov telescopes are, by design, also very efficient detectors of fast optical transients. So far not particularly stringent flux upper limits in the VHE $\gamma$-rays domain have been reported by the VERITAS, HESS and MAGIC telescopes.

Dedicated follow-up observations of FRB~20150418A were obtained with the HESS imaging atmospheric Cherenkov telescope array~\cite{HESS_Coll17}. HESS is sensitive to cosmic and $\gamma$-rays in the 100 GeV to 100 TeV energy range and is capable of detecting a Crab-like source close to zenith and under good observational conditions at the $5 \sigma$ level within less than one minute.
No high-energy afterglow emission was found in the 1.4 h of observational data for $E > 350$ GeV.
Assuming a distance of $z = 0.492$, the resulting 99\% C.L. upper limit on the $\gamma$-ray luminosity at 1 TeV is $L_{\rm VHE} < 5.1 \times 10^{47}$ erg s$^{-1}$.

As for other energy bands, FRB~20121102A also received particular attention by VHE observatories, in particular VERITAS~\cite{Bird2017} and MAGIC~\cite{MAGIC18}, conducting coordinated observations with Arecibo.
The details are reported in the Section \ref{sec:121102VHE}.
To note that because at $\sim$100 GeV the flux scales as $\sim$$d^2$, and by a larger factor at higher energies because of the reduced EBL attenuation, VHE observations simultaneous with radio bursts of closer repeaters, like FRB~20180916B, are potentially able to provide strong constraints on the magnetar models for FRB progenitors.

Finally FRBs spatially and temporally coincident with neutrino events in the TeV–PeV regime were searched by the ANTARES and the IceCube Collaborations.

As already mentioned, in the case of the Metzger et al. (2019)~\cite{Metzger19} synchrotron maser model, the upstream medium is considered to be a mildly relativistically expanding, baryon-loaded outflow with an electron-ion composition. Because both the radio and X-ray emission originate from the same physical location, their nearly simultaneous observed arrival time, like for FRB~20200428A, is naturally expected.
Another important consequence is the potential presence of ions with relativistic energies that can generate neutrino emission via the photohadronic interaction with thermal synchrotron photons, similar to proposed mechanisms of neutrino emission in $\gamma$-ray burst jets (see references in~\cite{Metzger20neutrino}).
%synchrotron maser electron-ion blast wave scenario
%
A burst of $\sim$TeV--PeV neutrinos of total energy $E_\nu \approx 10^{35-44}$~erg with a timescale $t_{\rm max} \sim 0.1$--1000~s following the radio burst is predicted~\cite{Metzger20neutrino}. Such neutrino detection looks extremely challenging with present detectors, and possibly even with future ones, and would only be possible for a giant flare from a nearby Galactic magnetar.

The ANTARES search considered 12 FRBs discovered in the period 2013--2017. Neutrino fluence upper limits were derived using a power-law spectrum and assuming spectral indexes $\gamma = 1.0, 2.0, 2.5$~\cite{Albert19}.
The non-detection was not a surprise as all the hadronic models considered in the paper predict signal from a single FRB source orders of magnitude below their detection threshold.
Alternatively, the non-detection of a neutrino signal from FRBs could be the result of non-hadronic production mechanisms in the FRB environment, or of the presence of a beamed jet of neutrinos.
On the other hand the authors estimate that the large rate of FRBs events over the entire sky could contribute to the neutrino diffuse emission, especially at energies $E_\nu < 60$ TeV, blazars being the other candidate sources.

A similar search for high-energy neutrinos from FRBs was performed by the IceCube Collaboration~\cite{Fahey17, Aartsen18, Aartsen20}.
Despite the larger detection volume and $\sim$10 (or more) higher sensitivity with respect to the ANTARES telescope, no significant signal was found.
To note also that while ANTARES location in the Mediterranean Sea makes it particularly sensitive to events occurring in the Southern sky, IceCube is sensitive mostly to FRBs occurring in the Northern hemisphere, where the derived upper limits on the neutrino fluence for a $E^{-2}$ spectrum are about a factor 20 more stringent than those determined by ANTARES at its maximum sensitivity.
In their study Aartsen et al. (2020)~\cite{Aartsen20} considered 28 one-off FRBs plus the repeater FRB~20121102A (39 events in total). All the one-off FRBs were searched for simultaneous tracklike events from muon neutrinos above $\sim$100 GeV whereas 9 of them and the 11 burst from FRB~20121102A were investigated for events in the MeV regime. These represent the first-ever limits on neutrino signals at MeV energies from FRBs. Upper limits on the time-integrated neutrino flux emitted by FRBs for a range of emission timescales from 10~ms to 1~day were set too.
As prospects for observation of an excess of MeV neutrinos in IceCube depends on the distance to the source, we expect interesting results for FRB~20180916B (and other similarly not too far sources) to be announced soon.

Prospects for the future are good. As the number of FRBs detected is quickly increasing, in particular in the North, we expected even more stringent upper limits in the near future.
Moreover in the next few years, combined analysis with the new generation KM3NeT/ARCA and the IceCube-Gen2 detectors will provide the most sensitive and homogeneous coverage of the neutrino sky ever reached for energies $E_\nu > 1$ TeV.
On the other hand, the collected data analysis would greatly benefit from more accurate models describing the neutrino production associated with FRBs to refine the constraints on the neutrino fluence and energy released.
%

%
%%%%%%%%%%%%%%%%%%%%%%%%%%%%%%%%%%%%%%%%%%
\section{FRB~20121102A and FRB~20180916B}
\label{sec:R1_R3}

FRB~20121102A and FRB~20180916B, the only two known objects showing periodic active phases ($161\pm 5$ and $16.35\pm 0.15$ days, respectively), and among the most active repeating sources, were the targets of several MWL campaigns/studies. We give here more in depth outcome of these observations.

%
%%%%%%%%%%%%%%%%%%%%%%%%%%%%%%%%%%%%%%%%%%
\subsection{FRB~20121102A}
FRB~20121102A is the most studied FRB source to date.
It was discovered in data acquired in 2012 at the Arecibo Observatory~\cite{Spitler14, Spitler16}.
In mid-2015, further observations revealed a series of bursts at a similar DM and sky position. This represented the discover of the fist repeating FRB and ruled out cataclysmic models for this source~\cite{Spitler16}.
Nine additional detections of FRB~20121102A were made with the VLA, leading to the its localisation with a precision of 0\farcs1 and the detection of a coincident persistent point-like source with an average flux density of 180~$\mu$Jy at 3~GHz and a $r \sim 25$ mag in Keck and GMOS images~\cite{Chatterjee17}. The source was not detected at 230 GHz with ALMA.

Deep optical and spectroscopic observations with the Gemini Observatory associated the FRB with a dwarf galaxy at $z = 0.19273$ (Figure~\ref{fig:rFRB_hosts}{a}), corresponding to a luminosity distances of 972~Mpc~\cite{Tendulkar17}.
The European VLBI Network detected four more bursts to localise the source with a precision of 0\farcs01 ($\sim$40 pc in linear distance)~\cite{Marcote17}, four orders of magnitude better than any other FRB.
As a consequence, it was the first FRB source to be targeted by various multiwavelength campaigns and archival searches, whose results in the X-/$\gamma$-ray and optical band are briefly reviewed in the following section.

%
%%%%%%%%%%%%%%%%%%%%%%%%%%%%%%%%%%%%%%%%%%
\subsubsection{X-/$\gamma$-ray Observations}
\begin{figure}[H]

\includegraphics[width=0.9\columnwidth]{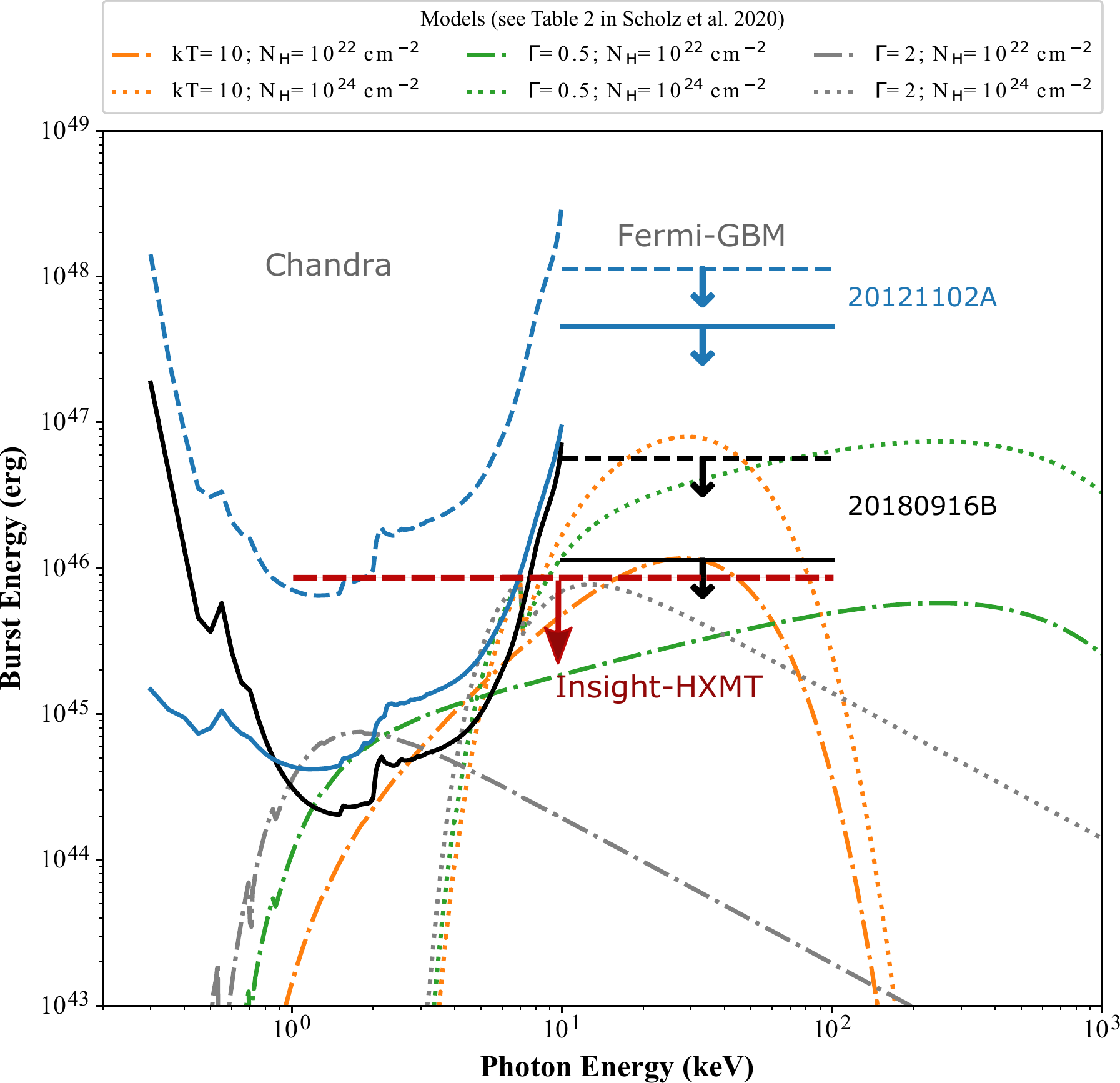}
\caption{Limits on energy of X-/$\gamma$-ray bursts at the time of radio bursts from FRB~20121102A (blue) and from FRB~20180916B (black) obtained respectively with {\em Chandra} ($0.5$--$10$~keV), with {\em Fermi}/GBM ($10$--$100$~keV), and with {\em Insight-HXMT} ($1$--$100$~keV). Dashed and solid lines are $5 \sigma$ upper limits derived on the integration time of a single FRB and stacked ones. {\em Chandra} and {\em Fermi} limits were derived simultaneously with FRBs, whereas no FRB was reported from FRB~20180916B for the {\em Insight-HXMT} one, which is given at 90\% confidence and assumes an integration time of $0.1$~s (from~\cite{Guidorzi20b}) (adapted with permission from Scholz, P., et al.; published by IOP Publishing, 2020~\cite{Scholz20}).
\label{fig:scholz20}}
\end{figure}

\textls[-15]{This source was first observed in X-rays with {\em Swift}/XRT ($0.3$--$10$~keV, 10 ks exposure time) and {\em Chandra} ($0.1$--$10$~keV, 39.5 ks) in November 2015 simultaneously with five radio facilities:  Arecibo, Robert C. Byrd Green Bank Telescope (GBT), Effelsberg, Lovell, VLA. A total of 17 bursts was detected, none covered by the X-ray observations and no credible persistent X-ray counterpart identified~\cite{Scholz16}. Scholz et al. (2017)~\cite{Scholz17} carried out another radio and X-ray campaign, this time in conjunction with {\em Chandra} and {\em XMM-Newton}, which led to the detection of radio bursts while the X-ray telescopes were observing. No X-ray signal was measured either simultaneous with the radio bursts or during the entire observation of 70~ks.
The $5\sigma$ upper limit for a persistent $0.5$--$10$~keV X-ray emission, assuming a photoelectrically absorbed power-law source spectrum, was $4 \times 10^{-15}$~erg~cm$^{-2}$~s$^{-1}$, equivalent to an isotropic-equivalent luminosity of $3 \times 10^{41}$~erg~s$^{-1}$ at the distance of FRB~20121102A~\cite{Scholz17} ($z = 0.19273$~\cite{Tendulkar17}, see also~\cite{Chatterjee17}).
Concerning the possible prompt X-ray counterpart to the radio burst, a stringer $5 \sigma$ upper limit was obtained of $3 \times 10^{-11}$~erg~cm$^{-2}$ in the $0.5$--$10$~keV energy band for a duration <$700$~ms, equivalent to an upper limit on the isotropic-equivalent released energy of $4 \times 10^{45}$~erg. The limits on fluence over 5-ms interval at any time during X-ray observations rise to $5 \times 10^{-10}$ and $10^{-9}$~erg~cm$^{-2}$. They also used the {\em Fermi}/GBM data to constrain the 10--100~keV fluence at the time of the radio bursts to <$4 \times 10^{-9}$~erg~cm$^{-2}$, equivalent to $E < 5 \times 10^{47}$~erg.
Figure~\ref{fig:scholz20} shows these limits (blue) compared with some emission models including black-body spectra with $kT = 10$~keV, power-laws with photon index $\Gamma = 2$, and cutoff power-law ($\Gamma = 0.5$; $E_{\rm cut} = 500$~keV) with different plausible photoelectric absorption values.}

An archival search for ms-long hard X-ray bursts in the 15--150~keV energy band was carried out in the {\em Swift}/BAT data from October 2016 to September 2017, providing a $5 \sigma$ upper limit of $10^{-7}$~erg~cm$^{-2}$ on 1-ms fluence, equivalent to an isotropic-equivalent energy of $10^{49}$~erg~\cite{Sun19}, which is well above the typical energy of the initial spike of the MGFs yet observed in the Galaxy.

\textls[-15]{Between 2017 September 5--11 {\em NuSTAR} observed FRB~20121102A in five separate intervals, covering one radio burst discovered with Effelsberg: for this radio burst, the corresponding $5 \sigma$ upper limit on the fluence in the 3--79~keV energy band was (6--40)~$\times 10^{-9}$~erg~cm$^{-2}$, that is, (0.6--5)~$\times 10^{47}$~erg of emitted energy~\cite{Cruces21}. No burst was detected instead, within the same observational campaign, when INTEGRAL was observing the source. Additional data exploitation are reported as underway (Gouiff\`es et al., in preparation).}

An archival search spanning eight-year time of {\em Fermi}/LAT data were used by Zhang and Zhang (2017) to derive an upper limit in the 100~MeV--10~GeV persistent emission possibly associated with FRB~20121102A of $4 \times 10^{-12}$~erg~cm$^{-2}$~s$^{-1}$, corresponding to an isotropic luminosity upper limit of $4 \times 10^{44}$~erg~s$^{-1}$~\cite{ZhangZhang17}. A power-law spectrum model with photon index $-2$ was assumed. A total of 18 time intervals were considered and the photons within $10^\circ$ around the FRB position were analyzed.
The limit becomes $\sim$$10^{45}$~erg~s$^{-1}$ for each of the single time-intervals.
Under the assumption that the FRB source is a newborn magnetar, these upper limits are not tight enough to pose significant constraints on the parameters space given by its magnetic field, spin period, and age. Similarly for other models that do not invoke a magnetar as the source of the repeating FRBs.

%
%
% Xiao \& Dai, 2017, ApJ, 846, 130 (constraints on magnetar activity from high-energy obs)

%
%%%%%%%%%%%%%%%%%%%%%%%%%%%%%%%%%%%%%%%%%%
\subsubsection{Optical Observations}
As mentioned above, initial Keck and Gemini observations led to the identification of a $r \sim 25$~mag point-like source at the position of FRB~20121102A~\cite{Chatterjee17}.
To better characterise the source, Tendulkar et al. (2017)~\cite{Tendulkar17} obtained a series of deep Sloan Digital Sky Survey (SDSS) %please define
%R: done
$r^\prime$, $i^\prime$, and $z^\prime$ filters images and nine 1800 s spectroscopic exposures using GMOS, %please define
%R: already defined in Section 4.2
which led to the association with a low-metallicity dwarf galaxy at $z = 0.19273 \pm 0.0008$ and estimated diameter of $\lesssim$4 kpc (Figure~\ref{fig:rFRB_hosts}{a}).
They also estimated a stellar mass of $M_* \sim (4$--$7) \times 10^7 M_\odot$, and a star formation rate of $(\simeq$$0.23 M_\odot$ yr$^{-1}$,
later revised to $0.15 \pm 0.04 M_\odot$ yr$^{-1}$ by~\cite{Heintz20}.
Recent {\em HST} observations gave a galaxy diameter of $\sim$2 kpc, a stellar mass of $(1.4 \pm 0.7) \times 10^8 M_\odot$ and AB magnitude $23.435 \pm 0.055$ (not corrected for Galactic extinction)~\cite{Mannings20}.
The association of FRB~20121102A with a compact, persistent radio source is consistent with bursts coming from a young magnetar that powers a luminous pulsar wind nebula~\cite{Kashiyama17}. At the same time, models that scale with stellar mass or star formation do not predict association with galaxies like that hosting FRB~20121102A~\cite{Nicholl17}. These galaxies are the preferred environment for LGRBs and hydrogen-poor SLSNe-I, which have been suggested are signatures of magnetar birth~\cite{Modjaz08, Lunnan14}.

The first search for an optical burst simultaneous to FRB~20121102A radio events was conducted with the ULTRASPEC camera mounted on the 2.4-m Thai National Telescope~\cite{Hardy17}.
During a monitoring campaign together with the 100-m Effelsberg Radio Telescope, 13 radio bursts were detected but no optical flash.
The camera was acquiring frames of 70.7 ms each during a total observation time of 19.6 h. The filter was a custom-made broad-band filter comprising the SDSS $i^\prime$ and $z^\prime$ with a central wavelength of 767~nm. The $5 \sigma$ upper limit to the optical burst fluence was of 0.046~Jy~ms. This limit was obtained stacking the two nearest frames closer to each of the 12 observed radio bursts, which resulted in flux density detection limit of 0.33~mJy, corresponding to an upper limit on optical radiation simultaneous with a burst $\nu L_\nu \lesssim 10^{47}$ erg~s$^{-1}$. Multiplying by the duration of two frames (141.4 ms) gives the quoted upper limit.
Additional Effelsberg, GBT %please define
%R: done in the previous section
and Arecibo monitoring campaigns in 2018 and 2019 were performed together with the GTC to search for simultaneous optical bursts using HiPERCAM~\cite{Cruces21}. No radio bursts were detected in the Effelsberg data (covering the GTC %please define
exposures) during the simultaneous optical observations, and, at our knowledge, no optical upper limits estimate have been published so far.

Karpov et al. (2019)~\cite{Karpov19} observed the field of FRB~20121102A for three nights in September 2017 with the 50-cm D50 telescope. Using a $2\times 2$ binning and either full-frame or central half-frame readout, 47 or 86 Hz frame rates were achieved. From the more than 2.5 millions of collected frames, upper limits of 10 to 20 mJy in $\simeq$10--20 ms were obtained, depending on the filter used.

VHE Cherenkov telescopes are also capable to provide, by construction, optical blue-band observations. While very rough in terms of spatial resolution, their sub-millisecond time sampling capabilities are only rivaled by the optical fast photometers discussed in Section \ref{sec:opt_instr}.
Unfortunately the optical configuration and large FoV make them prone to be contaminated by FOBs caused by sources like meteors, satellites, space debris and observatory close-by light flash producers (e.g., car lights).
These background contributors significantly reduce the confidence of a potential FRB detection, unless it is quasi-simultaneous with the radio event.
MAGIC used this capability to investigate the $U$-band light collected by its modified unit II camera central pixel~\cite{Hassan17}, which covers a 0\fdeg1 deg FoV, during a set of coordinated observations of FRB~20121102A with Arecibo~\cite{MAGIC18} (see also the following Section). No significant excess was detected simultaneously ($\sim$$\pm$ 100~ms) with any of the five FRBs detected during the monitoring campaign. Assuming a 1~ms long burst, the quoted sensitivity upper limit was 6.6~mJy and 4.1~mJy for a single burst or stacking the data around the five events, respectively. These limits, though at the opposite side of the optical band, are significantly more stringent than those reported by ULTRASPEC and imply a radio-to-optical flux density slope $\alpha \lesssim -0.32$.
However, a bright $\sim 29$~mJy, 12~ms (FWHM) long burst was detected 4.3~s before the arrival time of the first FRB.
Statistical considerations on the occurrence of similar events lead the MAGIC team to interpret it as a spurious background event.

\subsubsection{VHE $\gamma$-rays Observations}
\label{sec:121102VHE}
\textls[-15]{Though VHE emission is expected to be significantly attenuated by the interaction with the EBL and it is not expected to produce a detectable flux above 1 TeV ($\gamma$-ray opacity depends on the energy as well as on the source distance), VHE emission from FRB~20121102A in the energy region of $\sim$100 GeV will be largely unaffected by the EBL and could be detectable.
Vieyro et al. (2017)~\cite{Vieyro17} proposed a model for FRB~20121102A based upon an (nowadays obsolete) AGN scenario and concluded that high energy emission may be detectable for seconds to minutes after the radio burst, even for modest energy budgets.
Bird and VERITAS Collaboration (2017) constrained the persistent very-high-energy emission with VERITAS to upper limits of $5.2 \times 10^{-12}$ and $4.0 \times 10^{-11}$ cm$^{-2}$ s$^{-1}$ TeV$^{-1}$ at their energy thresholds of 200 and 150 GeV (assuming power-law spectra with indices $-2$ and $-4$, respectively). This corresponds approximately to three orders of magnitude above the VHE flux that would be expected assuming a Crab nebula like source. Of the effective 10.83 h of observation, about 6.5 included coincident radio observations with Arecibo. No bursts were reported, both in VHE and in the optical either in blind or targeted search mode.}

\textls[-15]{The MAGIC telescopes also performed simultaneous observations of FRB~20121102A with Arecibo during several epochs in 2016--2017~\cite{MAGIC18}. While in the radio band (central frequency of 1.38 GHz) five bursts were detected, no millisecond time-scale burst emission was detected in VHE $\gamma$-rays or the optical band.
Considering the whole 22 h of good quality observational data, an average integral luminosity upper limit $L_{\rm VHE} \lesssim 1 \times 10^{45}$ erg s$^{-1}$ was obtained for energies above 100 GeV at 95\% C.L. (limits above 1 TeV would be a factor $\sim$10 less constraining). If only the total duration of the five FRBs is considered, the upper limit becomes $L_{\rm VHE} \lesssim 1 \times 10^{49}$ erg s$^{-1}$. }

It is expected that further simultaneous observations of FRB~20121102A by existing VHE facilities and the forthcoming Cherenkov Telescope Array (CTA) will result in an increased sensitivity of at least one order of magnitude as the number of observed FRBs increases, in fact reaching the emission range of known MGFs~\cite{PopovPostnov13,Lyubarsky14}.

%
%%%%%%%%%%%%%%%%%%%%%%%%%%%%%%%%%%%%%%%%%%
\subsection{FRB~20180916B}
\label{sec:FRB1809}
FRB~20180916B~\cite{chime19_8repeaters} was discovered by CHIME that initially reported 10 bursts with a flux density in the range $\sim$0.4--4~Jy. Follow-up VLBI campaigns, favored by the active nature of the source and its low ${\rm DM}\sim 349$~pc~cm$^{-3}$, led to its precise localisation (0\farcs023) and the identification of the host galaxy at a redshift $z = 0.0337 \pm 0.0002$ (luminosity distance of $149.0 \pm 0.9$~Mpc) (Figure~\ref{fig:rFRB_hosts}{c}) ~\cite{Marcote20}: the closest extragalactic FRB source yet discovered. This if we exclude the uncertain identification of the host of FRB~20171020A at $z \simeq 0.00867$.
This precise localisation immediately showed a dichotomy with the case of the original repeater, with FRB~20180916B associated to a star-forming region within a nearby massive spiral galaxy, at odds with FRB~20121102A, which is hosted in a dwarf galaxy~\cite{Chatterjee17, Marcote17}. The subsequent continuous monitoring of FRB~20180916B by CHIME led to the first identification of a periodicity in the active phases of a rFRB~\cite{Chime20-180916p}. The bursts active phase repeats every $16.3$~days, with an active window phase of approx $\pm 2.6$~days around the midpoint of the window.

This so far unique combination immediately brought it to the spotlight of MWL campaigns. The possibility of planning in advance on observing during time windows in which the source is more likely to emit radio bursts, combined with the $\sim$6.5 times closer distance with respect to FRB~20121102A, which gives a $\sim$40$\times$ increased flux gain, propelled a number of broadband simultaneous observations. Another interesting feature of this FRB is its apparent strongly frequency-dependent activity, due to the narrow-band nature of its radio bursts, with higher frequency events preferably occurring at early times in the active phase window~\cite{Chawla20, Aggarwal20VLA, Chime20-180916p}.
Multiple bursts were also recently reported by Low Frequency Array (LOFAR) %please define
%R: done
in the 110--188~MHz band, the lowest frequency detections of any FRB to date. The burst activity in the active phase window appears clearly and systematically delayed towards lower frequencies by $\sim$3~days from 600~MHz to 150~MHz~\cite{Pleunis21}. The frequency drift phenomena also seems present. Moreover, the LOFAR non-detection of five CHIME detected bursts confirm the narrow-band nature of repeating FRBs.
In spite no burst was so far detected at frequencies higher than $\sim$2 GHz, this could be suggestive of an even wider bursting active phase dependence over the observing frequency, possibly extending to high energies.

%
%%%%%%%%%%%%%%%%%%%%%%%%%%%%%%%%%%%%%%%%%%
\subsubsection{X-/$\gamma$-ray Observations}
We report here some of the most constraining results so far obtained on the X-/$\gamma$-ray activity of this source.

{\em Chandra} observed FRB~20180916B with two 16-ks exposures during the active phase of the periodic activity, simultaneously with CHIME, Effelsberg, Deep Space Network radio telescopes~\cite{Scholz20}. {\em Chandra} covered only one CHIME radio burst (no bursts were reported by the other telescopes at frequencies in the range $\sim$1.2--8.5 GHz), providing a $5 \sigma$ fluence upper limit of $5 \times 10^{-10}$~erg~cm$^{-2}$ in the $0.5$--$10$~keV energy band. This corresponds to an isotropic-equivalent limit $E < 1.6 \times 10^{45}$~erg, valid for any burst duration as far as it were contained within a time range [$-446$~s, 4.7~h] around the detected FRB. For a 5~ms X-ray burst arriving at any other time during the Chandra observations the energy limit becomes $E < 4 \times 10^{45}$~erg (see~\cite{Scholz20} for the details).
%hr change as h, please confirm
%R: OK
Regarding a possible persistent X-ray source located within the $1^{\prime\prime}$ radius region centered on the position of
FRB~20180916B, the $5\sigma$ upper limit on the persistent 0.5--10~keV X-ray absorbed flux (assuming a power-law spectrum with photon index $\Gamma \sim 2$ and
$N_{\rm H} \sim 1 \times 10^{22}$~cm$^{-2}$) was $8 \times 10^{-15}$ erg~cm$^{-2}$~s$^{-1}$, or $L_{\rm iso} < 2 \times 10^{40}$~erg~s$^{-1}$ at the luminosity distance of the source.
Using {\em Fermi}/GBM data collected simultaneous with radio bursts from this source, a $5 \sigma$ upper limit of $9 \times 10^{-9}$~erg~cm$^{-2}$ on the 10--100~keV fluence was obtained, equivalent to $E < 3 \times 10^{46}$~erg~\cite{Scholz20}. The corresponding limits for the only covered radio burst as well as the stacked limit for the multiple bursts covered with {\em Fermi}/GBM are shown in black in Figure~\ref{fig:scholz20}, to be compared with the analogous results for FRB~20121102A.
Panessa et al. (2020)~\cite{Panessa20} place a $3 \sigma$ upper limit on the 28--80~keV $\gamma$-ray flux of $3.4 \times 10^{-8}$~cm$^{-2}$~s$^{-1}$ for 100-ms long bursts at any time during their INTEGRAL observations. This is very similar to the {\em Fermi}/GBM limit when scaled to 3 instead of $5 \sigma$.

\begin{figure}[H]

\includegraphics[width=0.75\columnwidth]{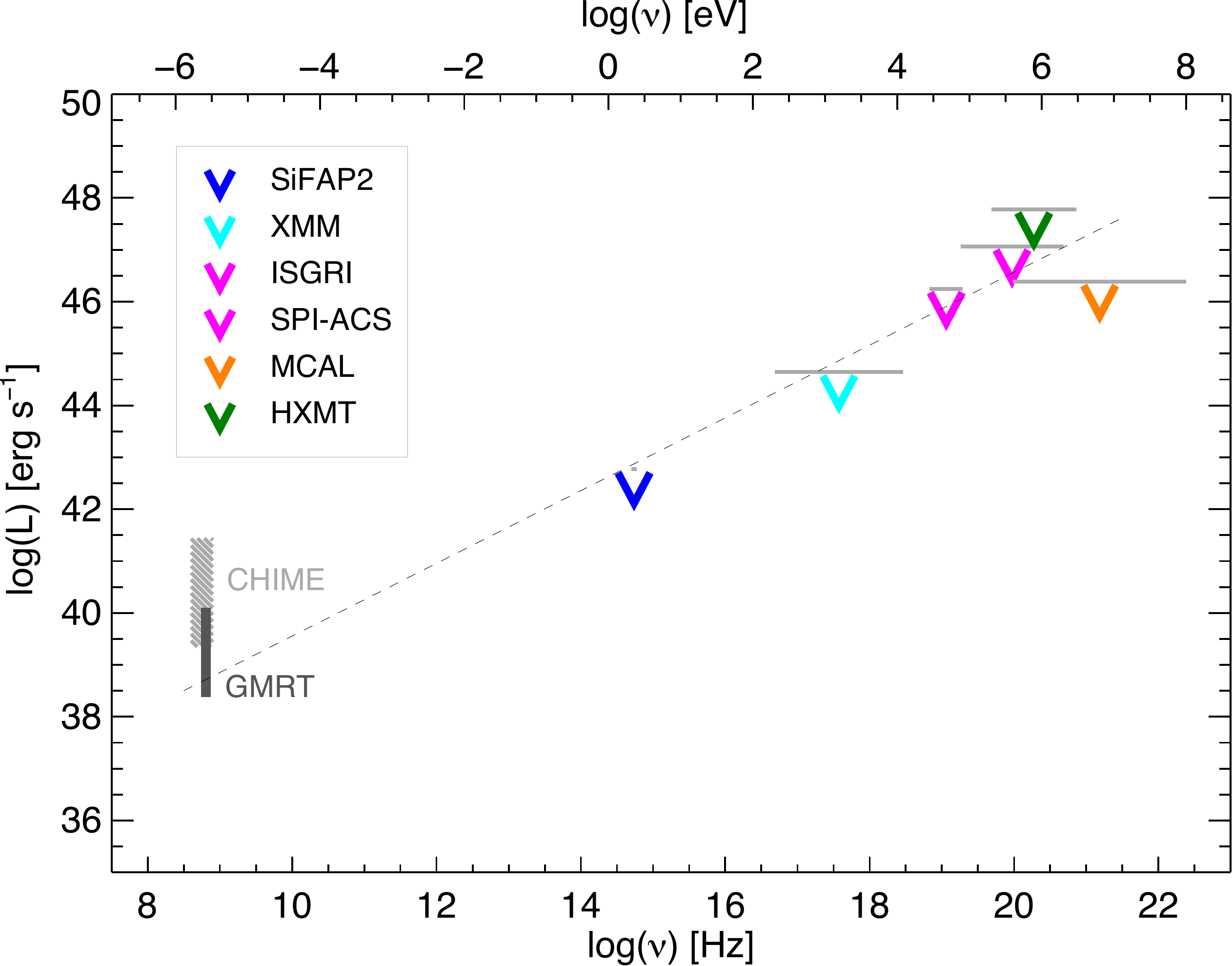}
\caption{FRB~20180916B burst luminosity vs frequency diagram throughout the electromagnetic spectrum.  The CHIME (400--800 MHz) and Giant Metrewave Radio Telescope (550--750 MHz) %please define the abbreviations
%R: CHIME defined in section 1 whereas GMRT has been replaced with its extended name
luminosity range are shown as grey areas (from~\cite{Chime20-180916p, Marthi20}). Adapted from the MWL campaign reported by~\cite{Pilia20}.
\label{fig:pilia20}}
%please confirm if change hyphgen as minu sign in the fire
%R: change done.
\end{figure}

Another successful case of broadband joint observations in which a radio burst was covered with a focusing sensitive X-ray telescope was reported by Pilia et al. (2020)~\cite{Pilia20}. Radio observations carried out with Sardinia Radio Telescope from 20 to 24 February 2020, corresponding to a peak in the periodic radio activity of the source, discovered three radio bursts at 328~MHz during the first hour and no one else during the following 30 h. In addition to a number of radio and optical facilities, also {\em XMM-Newton}, NICER, INTEGRAL, and AGILE joined the campaign, resulting in the X-ray coverage for all of the three bursts with {\em XMM-Newton} and for one with AGILE, with corresponding upper limits of $\sim$$10^{45}$~erg~s$^{-1}$ and of $3\sim10^{46}$~erg~s$^{-1}$ on the luminosity, respectively in the $0.3$--$10$~keV and MeV range during the bursts. The overall broadband upper limits as a function of frequency are shown in Figure~\ref{fig:pilia20}.

AGILE observations constrained the $3 \sigma$ level persistent average isotropic-equivalent luminosity in the MeV--GeV energy range to $L_\gamma (> 50$~MeV) $< (5$--$10) \times 10^{43}$~erg~s$^{-1}$ on a 100-day interval~\cite{Casentini20}.
In conjunction with radio observations, {\em Swift} and AGILE carried out a number of observations around the expected peaks of radio activity between February and March 2020, constraining the 0.3--10~keV luminosity $L_{\rm X} \lesssim 1.5 \times 10^{41}$~erg~s$^{-1}$~\cite{Tavani20}.
Additionally, a systematic search for $\gamma$-ray emission during 32 known radio bursts from this source did not reveal any coincidence on both millisecond and hours to days scales. Limits on the $>$$100$~MeV flux, that range from a few $10^{-7}$ to $\sim$$10^{-11}$~erg~cm$^{-2}$~s$^{-1}$, were obtained for integration times spanning from 10 to $\sim$$ 10^7$~s~\cite{Tavani20}. In terms of persistent flux, a long term integrated exposure gave an upper limit of $8.2 \times 10^{-13}$~erg~cm$^{-2}$~s$^{-1}$, corresponding to a luminosity $L_{\gamma} \sim 2 \times 10^{42}$~erg~s$^{-1}$.

{\em Insight-HXMT} observed the source around the peak expected between 4 and 7 February 2020 through a ToO observation. The unique combination of three instruments with a timing accuracy of $<$$10\; \mu$s, along with the unprecedentedly large geometric area in this energy range (5100~cm$^{2}$) makes it an optimal instrument to catch short faint bursts. Using a sophisticated set of triggering algorithms expressly devised and tailored to the instrumental background of each detector, upper limits on burst activity in the 1--100~keV energy band were obtained, constraining the released energy as a function of timescale, from 1~ms to 1~s: in particular, $E < 10^{46}$ erg for duration $\lesssim$$0.1$~s during several tens of ks exposure~\cite{Guidorzi20b} (see the black dashed thick limit in Figure~\ref{fig:scholz20}).

%
%%%%%%%%%%%%%%%%%%%%%%%%%%%%%%%%%%%%%%%%%%
\subsubsection{Optical Observations}
Untargeted optical observations of FRB~20180916B were obtained from the ZTF %please define
\cite{Andreoni20}. Images obtained during several nights for a total of 5.69 h set a median (Galactic extinction corrected) upper limit of $r \sim 18.1$ mag. This using images taken with 30 s exposure times.

On the other hand Pilia et al. (2020)~\cite{Pilia20} and Zampieri et al. (2020)~\cite{Zampieri20} reported high-speed optical observations with the IFI+Iqueye/Galileo, Aqueye+/Copernicus and SiFAP2/TNG camera/telescope during MWL observational campaigns. However no simultaneous optical coverage of the detected radio bursts was available. The most stringent 1-ms emission upper limit from SiFAP2/TNG was $V \sim 15.5$ (fluence $\lesssim 2$ mJy ms).

Recently Kilpatrick et al. (2020)~\cite{Kilpatrick20} reported about a deep optical search for afterglow emission with the Apache Point Observatory (APO) 3.5 m telescope.
The full sequence of $3 \times 30.33$ s, $gri$ images cover the CHIME detected post-burst epochs [+2.2, +1938.1] s, relative to the dispersion-corrected burst arrival time.
The obtained $3 \sigma$ limiting AB magnitudes were compared to predictions of the synchrotron maser model of Metzger et al. (2019)~\cite{Metzger19} and Margalit et al. (2020b)~\cite{Margalit20constr}.
As discussed above, one prediction of this model is that there should be a broadband (incoherent) synchrotron afterglow that will accompany and follow the FRB. On timescales similar to the FRB duration, this afterglow will peak in hard X-/$\gamma$-rays, but it can subsequently cascade through optical bands on timescales of minutes post-burst.

Considering FRB~20180916B typical values for (a) the circumburst densities of $n_{\rm ext} = 2000$ cm$^{-3}$~\cite{Margalit20constr}, (b) the ratio of electron to ion number densities in the upstream medium $f_e = 0.5$ and (c) a synchrotron maser efficiency $f_\xi \approx 10^{-3}$~\cite{PlotnikovSironi19}, an average energy per burst $E = 2.5 \times 10^{42}$ erg is obtained.
Moreover from the expected moment of dispersion corrected burst arrival at optical wavelengths, the timescale for the optical light curve is $t_{\rm syn} \approx 87$ s. By that time the optical luminosity is $\nu L_\nu \approx 6 \times 10^{38}$ erg s$^{-1}$ ($\approx$$28$ mag at 150 Mpc).
Such an optical counterpart would then be way below the range of detectability in the mentioned observations ($\approx$24.5 mag in individual $gri$ frames or $\approx$26 mag in the stacked frames).
It also shows that probing long time scales after the event is of little help, in the framework of the mentioned models.
On the other hand, if the burst profile is significantly more luminous on short timescales (one to tens of ms) after the burst, like for the models of {Beloborodov} (2017), high-speed cameras could probe detection threshold shallower by several orders of magnitude. Of course much different density profiles or shorter waiting time between bursts active periods, that would temporarily enhance the circumburst density, could lead to significantly different optical evolution, possibly with much longer duration afterglows.
Still high-speed cameras on large telescopes looks the most appropriate strategy to put the strongest constraints on potential optical counterparts.

The FRB~20180916B $\simeq 60$~pc close environment has been recently studied using the wide field camera 3 instrument on {\em HST} and the GTC/MEGARA integral field unit spectrograph~\cite{Tendulkar21}. These observations excluded the possibility of the presence of a satellite galaxy and showed that the FRB location is $250 \pm 190$~pc away from the nearest knot of active star formation. Among the various proposed possible progenitors, a neutron star high mass x-ray (HMXB) or $\gamma$-ray binary system with a late O-type or B-type companion seem to better explain the observed activity period, positional offset, and local emission.
High-cadence searches for bright radio bursts from Galactic HMXBs and $\gamma$-ray binaries can help to better establish a connection to FRB~20180916B.
To note also that a scenario in which periodic rFRBs are powered by transient flares from accreting stellar-mass BH or NS binary systems undergoing super-Eddington mass transfer, similar to those which characterize some ULX sources, have been recently proposed~\cite{Sridhar21}.

%
%%%%%%%%%%%%%%%%%%%%%%%%%%%%%%%%%%%%%%%%%%
\section{SGR~J1935+2154}
\label{sec:SGR1935}
The discovery of FRB~20200428A, a bright radio burst from the Galactic magnetar SGR~J1935+2154 (hereafter, SGR~J1935) on 28 April 2020 was a long sought-after turning point in the FRB--magnetar connection.
Discovered in July 2014 by {\em Swift} and soon afterward followed up with {\em Chandra} and {\em XMM-Newton} between 2014 and 2015, SGR~J1935 was found to be a magnetar with period $P=3.24$~s, with a characteristic age of $3.6$~kyr and an X-ray spindown luminosity of $2 \times 10^{34}$~erg~s$^{-1}$~\cite{Israel16}. It turned out to be a very active magnetar ever since, as it underwent major outbursts (periods characterised by an increase of the persistent X-ray luminosity up to $10^3$ times above the quiescent level, accompanied by bursting activity) in 2015 and 2016 and was one of the most prolific sources of bursts~\cite{Younes17, Lin20b}.
On 27 April 2020, SGR~J1935 entered a new active phase, characterised by an increase of the persistent X-ray luminosity by a factor of $\sim$50 within a few days~\cite{Borghese20}, accompanied by the emission of a forest of hundreds of magnetar short bursts~\cite{Palmer20, Younes20}.

On 28 April 2020, CHIME at 600~MHz and STARE2 at $1.4$~GHz detected from SGR~J1935 an extremely bright radio burst, FRB~20200428A, which consisted of two peaks 30-ms apart and with a fluence of $\sim$1.5~MJy~ms, as estimated by STARE2~\cite{Chime20SGR1915, Bochenek20Nat}. The corresponding released energy ranges from $3 \times 10^{34}$ to $2 \times 10^{35}$~erg (assuming a distance of $9.5$--$10$~kpc), corresponding to $\sim$$10^3$ times more energetic than any radio burst previously observed from magnetars and to just about one decade less energetic than the weakest extragalactic FRBs yet observed~\cite{Chime20SGR1915, Bochenek20Nat, Marthi20}. This lends strong support to the conjecture that active magnetars can be sources of extragalactic FRBs and that the energy distribution of FRBs likely extends down to comparably low values.

\begin{figure}[H]

\includegraphics[width=0.8\columnwidth]{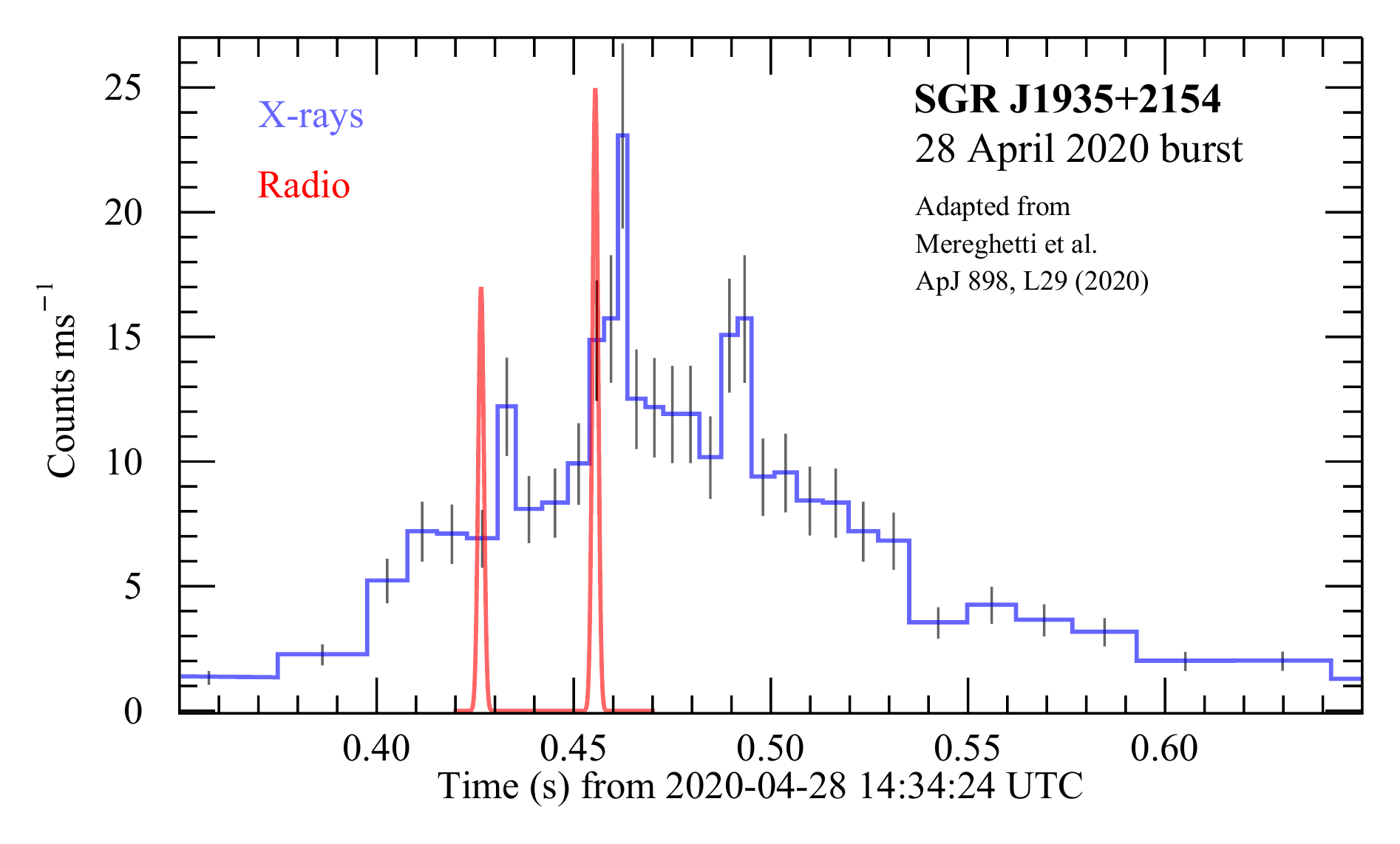}
\caption{INTEGRAL/IBIS-ISGRI time profile of the bright X-ray burst (20--200~keV) from Soft Gamma Repeater (SGR)~J1935 associated with FRB~20200428A. The red profile shows the FRB~20200428A radio pulses as observed with CHIME (adapted with permission from Mereghetti, S., et al.; published by IOP Publishing, 2020~\cite{Mereghetti20}).
\label{fig:mereghetti20_1}}
%please confirm if change hyphen as minus sign in the figure
%R: done
\end{figure}

{A bright simultaneous $\sim$1-s long X-ray burst was detected with {\em Insight-HXMT}~\cite{LiHXMT20}, which consisted of two major bumps $0.2$~s apart. The second bump, which was also much brighter than the first one, was also detected with INTEGRAL~\cite{Mereghetti20} and with {\em Konus}/WIND~\cite{Ridnaia20} and was characterised by three peaks $\sim$30~ms apart. Also AGILE detected it~\cite{Tavani21}. Once the delay due to the DM associated with the direction of SGR~J1935 is accounted for, the first two X-ray peaks temporally coincide with the two radio peaks of FRB~20200428A within a few ms~\cite{Mereghetti20, LiHXMT20, Ridnaia20} (Figure~\ref{fig:mereghetti20_1}). }
The X-ray spectrum of this burst can be modelled with a cutoff power-law with peak energy in the range $65$--$85$~keV and photon index $\Gamma=0.7$
\cite{Mereghetti20, Ridnaia20, LiHXMT20} with a fluence of $6.1 \times 10^{-7}$~erg~cm$^{-2}$ (20--200 keV~\cite{Mereghetti20}) and $7.1 \times 10^{-7}$~erg~cm$^{-2}$ (1--250 keV~\cite{LiHXMT20}), corresponding to a released energy ranging from $\sim$$10^{39}$ to $\sim$$10^{40}$~erg, depending on whether a distance of $4.4$ or $12$~kpc is assumed, respectively. While this burst is significantly harder than other events from this source and its time profile appears to be different from the bulk, its fluence is in line with the distribution~\cite{Mereghetti20, Lin20b, Ridnaia20}.

In the aftermath of the 2020 outburst, the persistent X-ray luminosity fading is described by the sum of two exponentials with very different e-folding times ($0.65\pm0.08$ and $75\pm5$~days), accompanied by the cooling of the black-body spectrum~\cite{Younes20b}.

FAST observed SGR~J1935 about 14~h prior to FRB~20200428A and covered 29 X-ray bursts detected with {\em Fermi}/GBM: no radio pulse was observed with upper limits as deep as $10^{-8}$ times the radio fluence of FRB~20200428A~\cite{Lin20}.
Nearly two days later, on 30 April 2020, FAST detected a highly polarised radio burst from SGR~J1935 with a fluence of $60$~mJy~ms~\cite{Zhang20_FAST1935}, with {\em Insight-HXMT} detecting no simultaneous X-ray burst down to $\sim$$10^{-8}$~erg~cm$^{-2}$ in the 1--250~keV band~\cite{Li20_HXMT_200430}. Next, a couple of bright radio bursts $1.4$~s apart from the same source were detected on 24 May 2020, with fluence of $112\pm 22$ and $24\pm 5$~Jy~ms~\cite{Kirsten20}.
%hr change as h, please confirm
% R: OK
Additionally, periodic radio pulsations with period $P \simeq 3.24760$~s was reported by a Northern Cross observation on 30 May 2020~\cite{Burgay20ATel}, which is 0.3~ms longer period than that measured by {\em NuStar}~\cite{Borghese20}. The derived average $\dot{P} \simeq 1.2\times 10^{-10}$~s~s$^{-1}$ is an order of magnitude larger than that derived from the X-ray analysis of the 2014 outburst~\cite{Israel16}.
Later, three more bright radio bursts were reported by CHIME on 8 October 2020, with fluence values of $900\pm 160$, $9.2\pm 1.6$, and $6.4\pm 1.1$~Jy~ms~\cite{Pleunis20}.

{While transient radio pulsations associated with magnetars in outburst were already known in a few cases~\cite{KaspiBeloborodov17, Lower20, Esposito20}, these observations show that SGR~J1935 can emit both radio and X-ray bursts independently of each other, with radio bursts spanning a fluence range of more than seven orders of magnitude, thus possibly narrowing the gap between magnetar sporadic radio burst emission and extragalactic FRBs.
The magnetar emission of sporadic bright radio bursts possibly associated with X-ray bursts is not unprecedented~\cite{Burgay18, Israel21}, although in no case other than SGR~J1935 the radio burst was so bright as to bridge the gap with extragalactic FRBs.
While SGR~J1935 makes a compelling case for a link between active magnetars and sources of extragalactic FRBs, the variety of properties exhibited by FRBs leaves the possibility that just some of them are due to extragalactic magnetar equally or even more active than SGR~J1935.}

\begin{figure}[H]

\includegraphics[width=0.5\columnwidth]{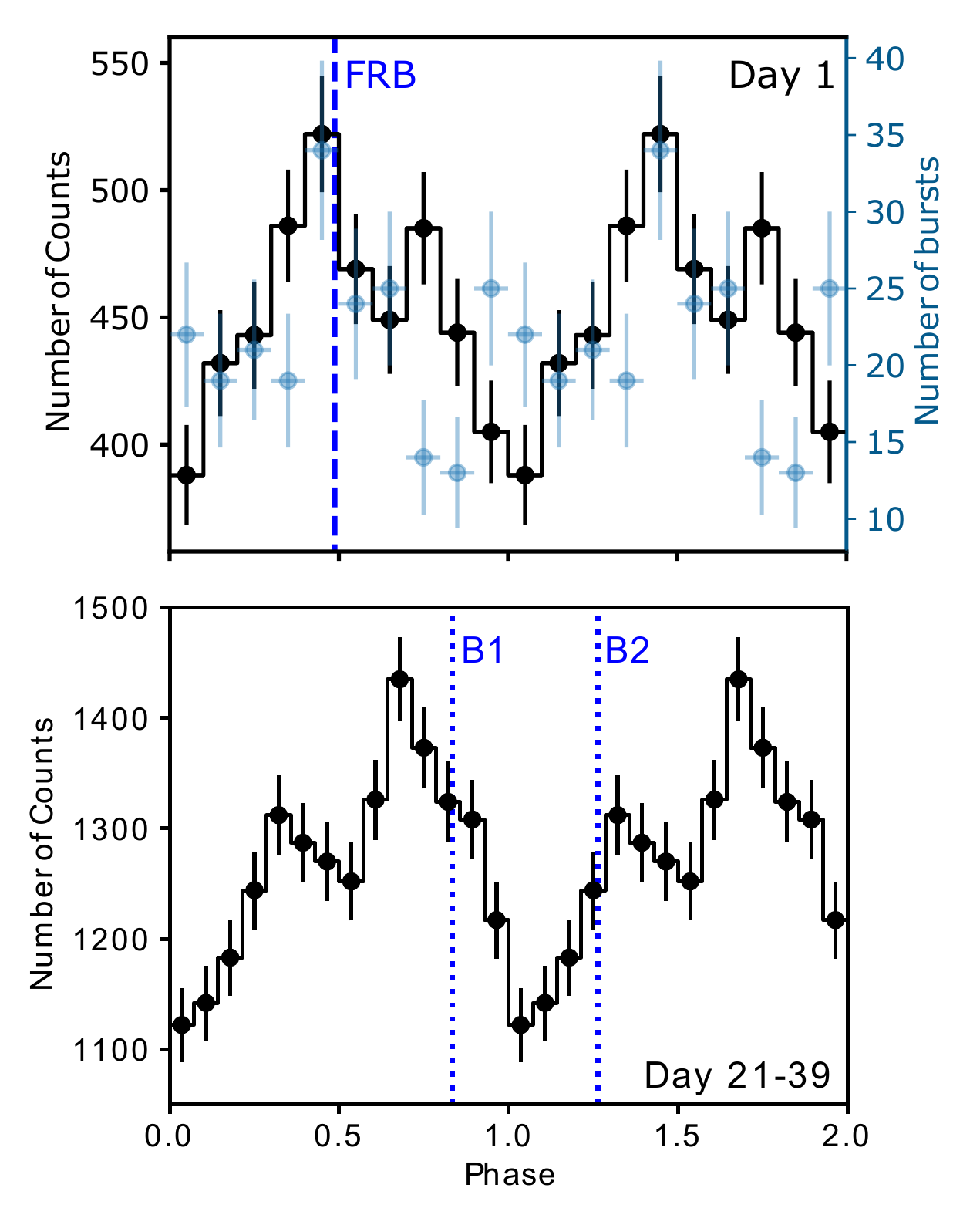}
\caption{({\textbf{Top}}): average pulse profile of SGR~J1935 measured with {\em NICER} in the 1.5--5~keV energy band during the first day following FRB~20200428A (black points) along with the peaks of the bursts folded at the spin period (light blue). The phase of FRB~20200428A is shown with the vertical dashed line. ({\textbf{Bottom}}): same average pulse profile in the time interval from 21 to 39 days post FRB. The blue vertical lines show the relative phase of the two radio bursts observed by {Kirsten} et al. (2020)~\cite{Kirsten20} (reproduced with permission from Younes, G. A., et al.; published by IOP Publishing, 2020~\cite{Younes20b}).
\label{fig:younes20_f5}}
\end{figure}

{\em NICER} observations in the 1.5--5~keV energy band were used to resolve the spin phase of both FRB~20200428A and of the two radio bursts emitted by SGR~J1935 about one month afterwards. As shown in Figure~\ref{fig:younes20_f5}, while FRB~20200428A aligns with the brighter X-ray peak of the double-peaked X-ray profile, the other two radio bursts appear to occur independently of the X-ray pulsed profile. While FRB~20200428A is likely to be causally connected with the simultaneous X-ray burst, these observations suggest that magnetar radio bursts have no spin-phase dependence as also found for magnetar 1E~1547.0$-$5408~\cite{Israel21}, in line with the behaviour of X-ray bursts~\cite{KaspiBeloborodov17}. The main peak of the pulse profile is commonly ascribed to the peak of a hot region of the NS surface as viewed by an observer (e.g.,~\cite{PernaGotthelf08}), possibly associated with the magnetic poles. Since FRB~20200428A is aligned with this peak, its observation could be interpreted with the magnetar being instantaneously viewed down the polar axis and, as such, connected with the polar magnetic field lines. The rarity of FRB~20200428A could be therefore explained as the result of the coincidence between the burst emission and the polar field lines aligned with the line of sight, whereas the other sporadic and much less bright radio bursts occur at different spin phases~\cite{Younes20b}.

Among the several mechanisms by which magnetars can emit FRBs, two classes are worth mentioning---in one case, the flaring magnetar emits a plasmoid of relativistic particles which then shocks the external medium at outer radii ($10^{14}$--$10^{16}$~cm) causing synchrotron maser emission which manifests itself as the FRB~\cite{Lyubarsky14, Beloborodov17, Beloborodov20, Metzger19, Margalit20constr}.
The relativistic beaming would explain the rarity of FRB emission compared with the high-energy one as well as the negligible time delay between radio and high-energy photons.

Alternatively, both hard X-ray and radio bursts could be the result of the sudden release of energy through magnetic reconnection in the magnetosphere close to the magnetar surface, with consequent plasma acceleration~\cite{Lyutikov02, Lyutikov03b, Kumar17}.
While both in principle can account for the radio and high-energy properties of SGR~J1935 bursts, some evidence favouring the magnetospheric origin over the relativistic shock interpretation has recently been put forward from the observations of the polarisation angle evolution across a number of radio bursts from the repeater FRB~20180301A~\cite{Luo20}. However, how wide-reaching the implications of this result are for the entire population of FRB sources and for the activity of magnetars like SGR~J1935, still remains an open issue that only future observations will be able to clarify.

Yang et al. (2021)~\cite{Yang21_1935comp} performed a detailed analysis of all the SGR~J1935 detected high-energy bursts, as detected by the various satellites, and compared their properties to those of the (only) FRB-associated X-ray burst.
The FRB-associated X-ray burst only distinguishes itself from the other bursts because of its non-thermal spectrum and higher spectral peak energy, but is otherwise ``normal''. A larger sample of FRBs detections/non-detections are needed to assess whether the FRB-associated bursts are truly atypical.
Comparison of the released energy of FRB~20200428A and its associated X-ray burst were used to compute the energy ratio $\eta = E_{\rm radio}/E_{X} = 2.9 \times 10^{-5} $~\cite{Bochenek20Nat, Li20_HXMT_200430}.
Assuming that such a ratio is typical for FRB--SGR burst associations, the authors compare the cumulative energy distribution of SGR bursts with that of radio bursts from FRB~20121102A in the C-band (4--8 GHz)~\cite{zhang18}. It is then expected that FRB related X-ray bursts should have energies in the range $E_{X} \sim 10^{42-44}$~erg.
Moreover, if FRB~20200428A and its associated X-ray burst are used as a calibrator, we expect to have one FRB event every $\sim$150 SGR bursts, so only a small fraction of magnetar bursts can produce FRBs, in agreement with the results obtained in~\cite{Lin20,Lu20}. A possible explanation, invoked by several authors, to explain missing FRBs in the majority of SGR bursts is again beaming. We consider this possibility highly probable.

Finally, if the claimed $\approx$251~days periodic windowed behavior in the times of burst detections for SGR~1935~\cite{grossan20_1935periodic} and the possible $398.20 \pm 25.45$~days periodicity found in the high-energy bursts from SGR~1806--20~\cite{Zhang21_1806periodic} will be confirmed, then the connection between magnetar short bursts and FRBs would be further strengthen.

%
%%%%%%%%%%%%%%%%%%%%%%%%%%%%%%%%%%%%%%%%%%
\section{Conclusions and Prospects}
\label{sec:conclusions}
The just announced discovery of FRB~20200120E with a DM of 87.82~pc~cm$^{-2}$, the lowest measured to date, demonstrates once again how the FRB research field is rapidly evolving~\cite{Bhardwaj21}. Its possible location on the outskirts of the nearby galaxy M81, at a distance of 3.6 Mpc, or in Milky Way halo, demonstrates that within the rapidly growing FRB sample we will soon be able to study several more close-by sources, and consequently significantly reduce the detection thresholds reported in this review. Still also for this source (41 times closer distance with respect to FRB~20180916B implies a 1700 fluence gain) all the X-ray searches were unfruitful and the large error region ($\approx$$3^\prime\times \,$1\farcm5) makes a cross identification in the optical/NIR band prohibitive.
This discovery represents a shining example of how hard can be the hunting for a FRB high-energy counterpart.
As in the case of other transient phenomena, for example, GRBs and Gravitational Wave events, in this review we have shown that the multi-wavelength and multi-messenger approach has the potential to solve several questions concerning FRB such as---Are all FRBs produced by a single astronomical class of objects? Are there any physical difference between one-off and repeating FRBs? Is repetition/periodicity a common feature hidden in a wide range of time scales? What are the physical processes involved? What is the MWL energetic involved? Can FRBs be used as cosmological probes?
In particular the long sought high-energy light bursts from a FRB source would mark a leap in the study and understanding of these intriguing phenomena.
Dedicated instrumentation, either existing or to be built, eventually shared with other transient sources hunting teams, is a key point. In the optical/NIR band, the usage of fast cameras/photometers looks like the most promising, if not the only strategy, though new general purposes observational resources, like the Vera Rubin telescope, could still be a valuable resource to test a number of FRB emission models.

From the several observational efforts and findings summarised in this review, and accepting as favoured models those involving a magnetar as central engine, we can draw the following conclusions:
\begin{enumerate}
\item MWL follow-up approaches like those implemented for GRB afterglows are likely to produce null results as delayed ($\gtrsim$1--10 s) high-energy emission is either unlike or too weak to be detected by present and future instruments;
\item searches for transients in MWL archives are meaningful only for periodic FRBs for which searches in an active phase window can give statistically significant upper limits or detections;
\item searches for FRBs host galaxies can be successful for nearby (based on the DM--$z$ relation) events or for small error areas (of the order of $5^{\prime\prime}$) and would take advantage of photometric or (better) spectroscopic information of the potential hosts;
\item \textls[-20]{ specific FRBs observational campaigns, like those of FRB~20121102A and FRB~20180916B, and MWL searches involving simultaneous sky coverage by radio and higher energy telescopes, capable of high frequency acquisition, represent the most promising strategy to detect or to set stringent upper limits of a bursts in an energy band other than the radio;}
\item a reversed new FRBs search strategy where a few small radio telescopes, like the 4.5-m diameter DSA-10/-110 dishes~\cite{Kocz19}, are employed to shadow the {\em Swift}/XRT, and other X-ray telescopes, pointings could represent a cheap and ready alternative to scheduled MWL campaigns. This would also be useful to test the recently proposed ULX binary scenario~\cite{Sridhar21};
\item the MeerLICHT approach, where an optical telescope with a relatively large FoV is co-pointing a radiotelescope/interferometer, is also very interesting, but to make it effective EMCCD or fast photometers must be employed to reach a time resolution as short as $\sim$1--10~ms. To note that in the case of searches of new FRBs for which a small sky/detector area cannot be selected, the data throughput would become prohibitive, so that a compromise on acquisition frequency/sky-area must be adopted;
\item MWL campaigns of Galactic magnetars are highly needed to test emission characteristics and fluence ranges in the framework of the proposed unified magnetar models;
\item the $\approx$1000 new FRBs that will be soon announced by the CHIME collaboration (preliminary results reported in a public seminar) and the other search programs will deliver new information that will surely help to address the MWL search efforts.
\end{enumerate}

%%%%%%%%%%%%%%%%%%%%%%%%%%%%%%%%%%%%%%%%%%
%\vspace{6pt}

%%%%%%%%%%%%%%%%%%%%%%%%%%%%%%%%%%%%%%%%%%
%% optional
%\supplementary{The following are available online at \linksupplementary{s1}, Figure S1: title, Table S1: title, Video S1: title.}

%%%%%%%%%%%%%%%%%%%%%%%%%%%%%%%%%%%%%%%%%%
\authorcontributions{All authors contributed to the manuscript writing, in particular: conceptualization, L.N. and C.G.; methodology,  L.N., C.G. and E.P.; software, C.G.; validation, M.T., E.P. and A.G.; software, C.G.;
formal analysis, L.N., C.G. and E.P.; investigation, L.N., C.G., L.Z. and E.P.; data curation, C.G. and E.P.; writing---original draft preparation, L.N. and E.P.; writing---review and editing, All authors contributed; visualization, L.N., C.G. and L.Z. All authors have read and agreed to the published version of the manuscript.
%\href{http://img.mdpi.org/data/contributor-role-instruction.pdf}{CRediT taxonomy} for the term explanation. Authorship must be limited to those who have contributed substantially to the work reported.
}

%%%%%%%%%%%%%%%%%%%%%%%%%%%%%%%%%%%%%%%%%%
\funding{This research was partially funded by the Italian Space Agency (ASI) and National Institute for Astrophysics (INAF) grant ASI-INAF I/037/12/0 and ASI-INAF n.2017-14-H.0, and by INAF grant ``Sostegno alla ricerca scientifica main streams dell'INAF'' (Presidential Decree 43/2018). M.T. is partially supported by the PRIN-INAF 2017 project “Towards the SKA and CTA era: discovery, localization, and physics of transient objects”.
}

%%%%%%%%%%%%%%%%%%%%%%%%%%%%%%%%%%%%%%%%%%
\acknowledgments{We would like to thank the two anonymous referees for their thorough and constructive report which helped to improving this review.
This research has made use of NASA’s Astrophysics Data System.
}
%

%\institutionalreview{Not applicable.}
%In this section, please add the Institutional Review Board Statement and approval number for studies involving humans or animals. Please note that the Editorial Office might ask you for further information. Please add “The study was conducted according to the guidelines of the Declaration of Helsinki, and approved by the Institutional Review Board (or Ethics Committee) of NAME OF INSTITUTE (protocol code XXX and date of approval).” OR “Ethical review and approval were waived for this study, due to REASON (please provide a detailed justification).” OR “Not applicable” for studies not involving humans or animals. You might also choose to ex-clude this statement if the study did not involve humans or animals.

%\informedconsent{Not applicable.}
%Any research article describing a study involving humans should contain this statement. Please add “Informed consent was obtained from all subjects involved in the study.” OR “Patient con-sent was waived due to REASON (please provide a detailed justification).” OR “Not applicable” for studies not involving humans. You might also choose to exclude this statement if the study did not involve humans. Written informed consent for publication must be obtained from participating patients who can be identified (including by the patients themselves). Please state “Written informed consent has been obtained from the patient(s) to publish this paper” if applicable.

\dataavailability{Data presented in this study whose source is not explicitly reported are available on request from the corresponding author.}
%In this section, please provide details regarding where data supporting reported results can be found, including links to publicly archived datasets analyzed or generated during the study. Please refer to suggested Data Availability Statements in section “MDPI Research Data Policies” at \href{https://www.mdpi.com/ethics}{https://www.mdpi.com/ethics}. You might choose to exclude this statement if the study did not report any data.

%%%%%%%%%%%%%%%%%%%%%%%%%%%%%%%%%%%%%%%%%%
\conflictsofinterest{The authors declare no conflict of interest.
The funders had no role in the design of the study; in the collection, analyses, or interpretation of data; in the writing of the manuscript, or in the decision to publish the results}
%
%\end{paracol}

%=====================================
% References, variant A: internal bibliography
%=====================================
\reftitle{References}

%=====================================
% References, variant B: external bibliography
%=====================================
%\externalbibliography{yes}
%\bibliography{biblio.bib}

%%%%%%%%%%%%%%%%%%%%%%%%%%%%%%%%%%%%%%%%%%
%% optional
%\sampleavailability{Samples of the compounds ...... are available from the authors.}

%% for journal Sci
%\reviewreports{\\
%Reviewer 1 comments and authors’ response\\
%Reviewer 2 comments and authors’ response\\
%Reviewer 3 comments and authors’ response
%}

%%%%%%%%%%%%%%%%%%%%%%%%%%%%%%%%%%%%%%%%%%
\end{document}